\documentclass[pra,twocolumn,showpacs,superscriptaddress,floatfix, nofootinbib]{revtex4-1}
\usepackage[caption=false]{subfig}
\usepackage{amsmath,graphicx}
\usepackage{epsfig}
\usepackage{amsmath,amssymb,mathrsfs,esint}
\usepackage{gensymb}
\usepackage{comment}
\usepackage[bookmarks=true,
   colorlinks=true,
   linkcolor=blue,
   urlcolor=blue,
   citecolor=blue,
   bookmarks=true,
   hyperindex=true
]{hyperref}
\usepackage{natbib}
\usepackage{relsize}
\usepackage{float}
\usepackage{dsfont}
\usepackage{bm}

\newcommand{\be}{\begin{equation}}
\newcommand{\ee}{\end{equation}}

\newcommand{\beq}{\begin{eqnarray}}
\newcommand{\eeq}{\end{eqnarray}}

\def\H1{\widehat{H}_1}

\newcommand{\ket}[1]{\left| #1 \right>}
\newcommand{\bra}[1]{\left< #1 \right|}

\newcommand{\bs}{\boldsymbol}

\begin{document}

\title{Floquet integrability and long-range entanglement generation \\ in the one-dimensional quantum Potts model
}

\author{A.\,I. Lotkov}
\affiliation{Russian Quantum Center, Skolkovo, Moscow 143025, Russia}
\affiliation{Alikhanov Institute for Theoretical and Experimental Physics NRC ``Kurchatov Institute'', Moscow 117218, Russia }

\author{V. Gritsev}
\affiliation{Institute for Theoretical Physics Amsterdam, Universiteit van Amsterdam, Amsterdam 1098 XH, The Netherlands}
\affiliation{Russian Quantum Center, Skolkovo, Moscow 143025, Russia}

\author{A.\,K. Fedorov}
\affiliation{Russian Quantum Center, Skolkovo, Moscow 143025, Russia}
\affiliation{Schaffhausen Institute of Technology, Schaffhausen 8200, Switzerland}
\affiliation{National University of Science and Technology ``MISIS'', 119049 Moscow, Russia}

\author{D.\,V. Kurlov}
\affiliation{Russian Quantum Center, Skolkovo, Moscow 143025, Russia}

\begin{abstract}
We develop a Floquet protocol for long-range entanglement generation in the one-dimensional quantum Potts model, which generalizes the transverse-filed Ising model by allowing each spin to have $n>2$ states. We focus on the case of $n=3$, so that the model describes a chain of qutrits.
The suggested protocol creates qutrit Bell-like pairs with non-local long-range entanglement that spans over the entire chain. 
We then conjecture that the proposed Floquet protocol is integrable and explicitly construct a few first non-trivial conserved quantities that commute with the stroboscopic evolution operator. 
Our analysis of the Floquet integrability relies on the deep connection between the quantum Potts model and a much broader class of models described by the Temperley-Lieb algebra. 
We work at the purely algebraic level and our results on Floquet integrability are valid for any representation of the Temperley-Lieb algebra.
We expect that our findings can be probed with present experimental facilities using Rydberg programmable quantum simulators and can find various applications in quantum technologies.
\end{abstract}

\maketitle

\section{Introduction}

Over the past few years, a tremendous progress in the development of programmable quantum simulators of various nature has greatly pushed the research field in the direction of probing novel non-conventional states of quantum matter~\cite{Lukin2017,Monroe2017,Martinis2018,Blatt2018,Trotzky2012,Mazurenko2017,Lukin2019}.
Recent achievements include investigations of exotic non-equilibrium many-body states and phase transitions in strongly-correlated quantum systems, e.g. in quantum spin chains, such as the transverse-field Ising model (TFIM)~\cite{Lukin2017,Monroe2017} and its extensions~\cite{Lukin2019}. 
Specifically, the use of a Rydberg programmable simulator enables one to study generalizations of TFIM, in which each spin has $n>2$ states~\cite{Lukin2019,Sadchev2018}.
The $n$-state model with $\mathbb{Z}_n$ symmetry is known as the chiral clock model (CCM), whereas a model with a larger~($S_n$) symmetry comes under the name of the $n$-state Potts model. 
Both models exhibit incredibly rich physics and have been extensively studied in the context of quantum phase transitions~\cite{Lukin2017,Lukin2019,Sadchev2018}, 
critical phenomena~\cite{Lukin2019}, exotic quasi-particle excitations (e.g. mesonic and baryonic)~\cite{Gorshkov2020}, and integrable lattice models~\cite{integrability}. 
An additional interest to the $n$-state Potts model is motivated by the fact that it admits a description in terms of parafermions,  particles obeying non-trivial quasi-local anyonic statistics, which is linked to topological quantum computing~\cite{topological, Fendley2016, HutterLoss2016}. On the other hand, the~$n$-state model corresponds to an array of qudits, which are promising for improving the performance of various quantum computational schemes and algorithms, e.g. by using them in the multiqubit gate decomposition~\cite{White2009, Wallraff2012, Kiktenko2020, Gokhale2019}. 
In order to maximize the improvement, one has to use qudits with a certain number of internal states that depends on the spatial topology and connectivity of a quantum system~\cite{Kiktenko2020}. 
For example, in the case of a one-dimensional (1D) chain with all-to-all connectivity, the best performance is shown by qutrits (qudits with $n=3$ internal states)~\cite{Kiktenko2020}.

One of the key challenges that arise in controllable spin chains and generic many-body systems is the generation of entanglement, which is a crucial resource for applications in quantum computing, simulation, and metrology. 
The case of long-range entanglement is traditionally of special interest, since it plays a key role in the understanding of various many-body phenomena, with the paradigmatic example being the quantum magnetism~\cite{Anderson1987}. 
A powerful tool for generating states with a long-range entanglement and other non-trivial properties is provided by the periodic (Floquet) driving, 
which allows steering the dynamics to the desired state by a sequence of discrete time steps~\cite{Gritsev2008, Gritsev2017, Lorenzo2017, Bertini2019a, Bertini2019b, Bertini2020a, Bertini2020b, Fan2020, Klobas2020, Maskara2021}. 
Recently, this method has been used in the realization of exotic non-equilibrium quantum many-body states, such as discrete time crystals~\cite{Maskara2021} and quantum many-body scars~\cite{Maskara2021, Bluvstein2021}.
A periodic driving protocol for on-demand generation of long-range entanglement has been suggested for
a system of  
ultracold atoms in optical superlattices, a setup which simulates the one-dimensional (1D) spin-$1/2$ Heisenberg model with time-dependent exchange interaction~\cite{Gritsev2008}. In this system, the consecutive switching of the interaction between the spins on even and odd links of the chain transforms the initial short-range entanglement between the the nearest-neighbour spins into the non-local one.  
A similar protocol was later studied for the case of the 1D TFIM, where the non-local entanglement between the pairs of distant spins is generated by repeatedly switching the transverse field on and off~\cite{Lakshminarayan2005,Mishra2015,Pal2018,Naik2019}. 
A natural question is whether these long-range entanglement protocols can be extended beyond the spin-$1/2$ chains, such as the Heisenberg or Ising model, to the case of~${\mathbb Z}_n$ chains, e.g. $n$-state Potts model. This is of practical interest due to the aforementioned advantages for quantum computing that are offered by the qudit-based platforms.   

From the fundamental point of view, it is important to emphasize that despite a significant amount of known Floquet protocols generating various non-trivial quantum states, all these cases are rather exceptional and require the system to be fine tuned. In contrast, a generic interacting many-body quantum system subject to a periodic drive simply reaches an infinite temperature state. This happens besause the energy is continuously pumped into the system and in the general case there are no conservation laws that can prevent the system from heating~\cite{Lazarides2014, DAlessio2014, Ponte2015}. Thus, the situation is analogous to the phenomenon of termalization in statistical physics, which is characteristic of non-integrable systems in the absence of disorder~\cite{Srednicki1994, Rigol2016}. On the other hand, it is well known from statistical physics that integrable or localized many-body systems do not termalize due to a large number of conserved quantities (charges). Continuing the analogy between statistical models and systems with periodic driving, it is natural to expect that Floquet systems which do~{\it not} heat up to an infinite temperature must be also in a certain sense integrable and possess an extensive number of conservations laws. This is indeed the case, and the field of Floquet integrability is a growing research area (for a review of recent results see e.g.~\cite{Gritsev2017} and references therein), but a complete understanding is still missing. In particular, explicit construction of conservation laws for integrable Floquet protocols remains an open question. 

In this paper, we propose a Floquet protocol for iterative generation of non-local entangled qutrit pairs in the~1D~$3$-state Potts model, which describes a chain of qutrits. We show that by starting from a polarized state (i.e., the product state in which all qutrits are initially in one and the same internal state) and performing a state preparation scheme followed by a consequent switching of the transverse field on and off with a certain frequency, one arrives at a state consisting of qutrit pairs with increasingly long-ranged entanglement. 

We then go one step further and argue that the existence of the suggested Floquet protocol is not merely a fortunate coincidence, but a consequence of its integrability. Namely, we demonstrate the presence of emerging conservations laws in the parameter regime corresponding to the long-range entanglement generation. We explicitly construct the first few conserved charges and conjecture that one can similarly construct the higher ones. Using the fact that the Hamiltonian of the $3$-state Potts model can be thought of as a representation of a more general operator belonging to the so-called Temperley-Lieb algebra, we show that the long-range entanglement generating Floquet protocols for the TFIM and the Heisenberg models are also integrable as their Hamiltonians are nothing other as different representations of the same Temperley-Lieb-algebraic operator.
Finally, we briefly discuss different driving regimes that do not result in the entanglement generation but  nevertheless exhibit some interesting features, although their detailed investigation is beyond the scope of the present work.

The rest of the paper is organized as follows. 
In Section~\ref{sec:Hamiltonian} we introduce the 1D 3-state Potts model and discuss an operator basis convenient for our purposes.  
Then, in Section~\ref{sec:Floquet_protocol} we construct the Floquet protocol, identify the parameter regime that leads to the entanglement generation, and present the resulting many-body state with the long-range entanglement between the qutrit pairs.  In Section~\ref{sec:Floquet_integrability} we show that the suggested Floquet protocol is integrable, present a few first non-trivial conserved charges, and extend our findings to other protocols related to the Temperley-Lieb algebraic models.  
We discuss our results and conclude in Section~\ref{sec:Conclusion}.

\section{3-state Potts model Hamiltonian}\label{sec:Hamiltonian}

We consider the Potts model with $n=3$ states per site, a generalization of the Ising model to spin variables taking three values. 
The Hamiltonian of the 3-state Potts model on a chain of $2N$ sites 
can be written as 
\be \label{Potts_H_tot}
	H = -J H_1 -f  H_2,
\ee
where $J$ and $f$ are real constants and for later convenience we separated the terms~$H_1$ and~$H_2$, which are given by
\be \label{Potts_H1_H2_XZ}
\begin{aligned}
	H_1 = &  \sum_{j=1}^{2N-1}\left(\, X_j^{\dag} X_{j+1} + X_j X_{j+1}^{\dag} \right),\\
	H_2 = & \sum_{j=1}^{2N}\left(\,Z_j  + Z_j^{\dag} \right).
\end{aligned}
\ee
For concreteness, throughout this work we assume open boundary conditions although most of the results can be straightforwardly generalized to the periodic ones~\cite{boundaries_note}.
The operators $X_j$ and $Z_j$ in Eq.~(\ref{Potts_H1_H2_XZ}) satisfy the following relations:
\be \label{Z_X_props}
\begin{aligned}
	&X_j^3 = 1,  &&Z_j^3 = 1, \\
	&X_j^2 = X_j^{\dag} = X_j^{-1},  &&Z_j^2 = Z_j^{\dag} = Z_j^{-1}, \\
	&X_j  Z_j = \omega  Z_j X_j,   &&X_j Z_k  = Z_k X_j   \;(j\neq k),
\end{aligned}
\ee
where $\omega = e^{2\pi i/3}$ is the principal cube root of unity.  They act non-trivially on the $j$-th site of the chain, i.e. $X_j = {\mathds 1} \otimes \ldots {\mathds 1}\otimes X \otimes {\mathds 1} \ldots \otimes {\mathds 1} $ and $Z_j = {\mathds 1} \otimes \ldots {\mathds 1}\otimes Z \otimes {\mathds 1} \ldots \otimes {\mathds 1} $. For the operators~$X$ and~$Z$ we choose the following matrix representations: 
\be \label{Z_X_matrix_rep}
	Z= \begin{pmatrix} 
	      1 & 0 & 0 \\
	      0 & \omega & 0 \\
	      0& 0 & \omega^2 \\
 	   \end{pmatrix}, \qquad
	   X =  \begin{pmatrix} 
	      0 & 1 & 0 \\
	      0 & 0 &1 \\
	      1& 0 & 0 \\
 	   \end{pmatrix}.
\ee 
Labelling the local basis states as $\ket{l}_j$, with $l \in \{0,1,2\}$, we have $Z_j^m \ket{l}_j = \omega^{l m} \ket{l}_j$ and $X_j^m \ket{l}_j = \ket{l - m \mod 3}_j$, where~$m=1,2$.
The matrices $Z_j$ and $X_j$ generalize the Pauli matrices $\sigma_j^z$ and $\sigma_j^x$, correspondingly, and are commonly referd to as the shift~($X_j$) and clock~($Z_j$) matrices. 

In Eq.~(\ref{Potts_H1_H2_XZ}), the term~$H_1$ corresponds to the nearest-neighbour interaction between the spins, whereas the term~$H_2$ plays the role of the transverse field. 
The operators~$X_j$, $Z_j$, and their conjugates are related to each other by a unitary transformation $W_j = {\mathds 1} \otimes \ldots {\mathds 1}\otimes W \otimes {\mathds 1} \ldots \otimes {\mathds 1}$, with
\be
	W = \frac{1}{\sqrt{3}} \begin{pmatrix} 
	      1 & 1 & 1 \\
	      1 & \omega^2 & \omega \\
	      1 & \omega & \omega^2 \\
 	   \end{pmatrix},
\ee
which acts on the operators~$X_j$ and~$Z_j$ as
\be \label{X_Z_transfrom}
\begin{aligned}
	& W_j X_j W_j^{\dag} = Z_j, &&W_j Z_j W_j^{\dag} = X_j^{\dag},  \\
	&W_j X_j^{\dag} W_j^{\dag} = Z_j^{\dag}, && W_j Z_j^{\dag} W_j^{\dag} = X_j.
\end{aligned}
\ee
Thus, the transformation~$\prod_{j=1}^{2N} W_j$ applied to $H_{1,2}$ from Eq.~(\ref{Potts_H1_H2_XZ}) simply replaces~$X_j \leftrightarrow Z_j$. 

Unlike the Pauli matrices, $Z_j$ and $X_j$ alone do not form a group under multiplication~\cite{gen_Pauli_Y_op}, which makes them inconvenient for our purposes. 
Thus, we choose a different basis that satisfies the group properties. 
Namely, following Refs.~\cite{Patera87, Fairliea89}, we introduce the operators~\cite{J_ops_definition_note}
\be \label{Z_X_to_J}
{\cal J}_j^{\boldsymbol{m}} \equiv {\cal J}_j^{(m_1, m_2)}  = \omega^{m_1 m_2} X_j^{m_1} Z_j^{m_2} ,
\ee
with $0\leq m_{1,2} \leq2$.  
Using the representation~(\ref{Z_X_matrix_rep}), from Eq.~(\ref{Z_X_to_J}) we obtain a unit matrix and eight unitary traceless matrices spanning the Lie algebra $\mathfrak{sl}(3,{\mathbb C})$. 
Taking into account that $X_j^{m_1} Z_j^{m_2}  = \omega^{m_1 m_2}  Z_j^{m_2} X_j^{m_1}$ and $\omega^{2 m_2 n_1} = \omega^{-m_2 n_1}$, one can easily show that the operators~(\ref{Z_X_to_J}) satisfy
\be \label{J_J_product}
	{\cal J}_j^{\boldsymbol{m}} {\cal J}_j^{\boldsymbol{n}} = \omega^{- \boldsymbol{m} \times \boldsymbol{n}} {\cal J}_j^{\boldsymbol{m} + \boldsymbol{n}},
\ee
where $\boldsymbol{m} \times \boldsymbol{n} \equiv m_1 n_2 - m_2 n_1$ and components of the vector $\boldsymbol{m} + \boldsymbol{n}$ are $\!\!\mod 3$. This leads to the following commutation relations:
\be \label{J_comm_rel}
	\left[  {\cal J}_j^{\boldsymbol{m}}, {\cal J}_k^{\boldsymbol{n}} \right] = -2i \, \delta_{jk} \, \sin \Bigl( \frac{2\pi}{3} \, \boldsymbol{m} \times \boldsymbol{n} \Bigr) {\cal J}_j^{\boldsymbol{m} + \boldsymbol{n}},
\ee
with $\delta_{jk}$ being the Kronecker symbol. 
In terms of the operators in Eq.~(\ref{Z_X_to_J}) we have
\be \label{Potts_H1_H2_J}
\begin{aligned}
	H_1 & =  \sum_{j=1}^{2N-1} \left( {\cal J}_j^{(2,0)}{\cal J}_{j+1}^{(1,0)} +  {\cal J}_j^{(1,0)}{\cal J}_{j+1}^{(2,0)}\right),\\
	H_2 & = \sum_{j=1}^{2N} \left( {\cal J}_j^{(0,1)} + {\cal J}_j^{(0,2)} \right). 
\end{aligned}
\ee
Note that all terms in $H_1$ ($H_2$) commute with each other, whereas $\left[ H_1, H_2 \right] \neq 0$. 
Let us also mention that the Hamiltonian~(\ref{Potts_H_tot}) is integrable at the critical point~$J=f$~\cite{integrability}, and its superintegrable variations are known~\cite{superintegrability}.
We now proceed with the discussion of a periodic driving protocol. 

\section{Floquet protocol}\label{sec:Floquet_protocol}

\subsection{Preliminary remarks}

We begin with a brief overview of the Floquet formalism for (isolated) time-dependent quantum systems. For a more detailed discussion see, e.g. Ref.~\cite{Bukov2015}.
The evolution operator for a time-dependent Hamiltonian $H(t)$ is given by the time-ordered exponential
\be \label{T_exp}
	U(t) = {\mathds T} e^{- i \int_0^{t} \, d\tau H(\tau)},
\ee
where ${\mathds T}$ denotes the time ordering and we set $\hbar = 1$. According to the Floquet theorem, for periodic time dependence of the Hamiltonian, $H(t+T) = H(t)$, one can rewrite Eq.~(\ref{T_exp}) in the following way:
\be \label{Floquet_unitary_single_exp}
	U_F(t) = P(t) e^{-i t H_F}, 
\ee
where $H_F$ is time-independent effective (Floquet) Hamiltonian, whereas the operator~$P(t)$ is periodic,  $P(t+T) = P(t)$, and satisfies $P(mT) = {\mathds 1}$ for integer $m$. Thus, if one observes the system {\it stroboscopically}, i.e. only at times $t = n T$ with integer~$n$,  the evolution operator becomes
\be \label{stroboscopic_Floquet_unitary_single_exp}
	U_F(T) = e^{-i T H_F}.
\ee
We note in passing that the Floquet Hamiltonian~$H_F$ can depend on the period duration~$T$. Despite the simple form of~Eq.~(\ref{stroboscopic_Floquet_unitary_single_exp}), explicit construction of~$H_F$ remains in most cases extremely tedious, if not impossible. Remarkable exceptions are provided by the Lie-algebraic~\cite{Gritsev2017} and free-fermionic~\cite{Arze2020} Hamiltonians, for which one can obtain~$H_F$ quite easily. 

An important and widely used class of periodic Hamiltonians corresponds to the so-called kicked models. A typical Hamiltonian is of the form
\be \label{H_kicked_general}
	H(t) = g_1 H_1 + g_2 T \sum_{m \in {\mathbb Z}} \delta(t - mT) H_2,
\ee
which describes a sequence of instantaneous kicks by the term~$g_2 H_2$ performed with a frequency~$\omega = 2\pi/T$. In Eq.~(\ref{H_kicked_general}) we explicitly include the factor of~$T$ in the second term in order to fix the dimension of the Hamiltonian. Substituting~$H(t)$ into Eq.~(\ref{T_exp}) with~$t=T$ and taking into account the~$\delta$-functional time dependence, we immediately obtain that the stroboscopic evolution operator factorizes and can be written as
\be \label{step-like_protocol_general}
	U_F(T) = e^{-i T g_2 H_2} e^{-i T g_1 H_1}.
\ee
Thus, over the period~$T$ the evolution is governed solely by~$g_1 H_1$, which is followed by the kick with~$g_2 H_2$ in the end. 

Alternatively, the stroboscopic Floquet protocol~(\ref{step-like_protocol_general}) can be obtained for the periodic step-like time dependence of the Hamiltonian:
\be \label{step_like_Hamiltonian_gen}
	H(t) = \begin{cases}
		H_1, \quad t \text{ mod } T_1 + T_2 \in [0, T_1), \\
		H_2, \quad t \text{ mod } T_1 + T_2 \in [ T_1, T_1 + T_2),
	\end{cases}
\ee
where~$H_{1}$ and~$H_2$ are time-independent. We thus have $H(t + T_1 + T_2) = H(t)$ and the one-period stroboscopic Floquet operator is
\be \label{step-like_protocol_general_2}
	U_F(T_1 + T_2) = e^{- i T_2 H_2} e^{- i T_1 H_1}.
\ee
Clearly, if $H_1$ and~$H_2$ do not commute, it is highly non-trivial to obtain the Floquet Hamiltonian~$H_F$ in a closed form. Indeed, in order to reduce the evolution operator from Eq.~(\ref{step-like_protocol_general}) or Eq.~(\ref{step-like_protocol_general_2}) to a single exponential as in~Eq~(\ref{Floquet_unitary_single_exp}), one has to sum the Baker-Campbell-Hausdorff series, which is only possible in a limited number of cases. Nevertheless, the form of the stroboscopic Floquet operator in Eq.~(\ref{step-like_protocol_general}) is already simple enough to work with and it has been investigated for various models and settings, see e.g.~\cite{Lakshminarayan2005, Mishra2015, Pal2018, Naik2019, Gritsev2008, Gritsev2017, Bertini2019a, Bertini2019b, Bertini2020a, Bertini2020b, Fan2020, Klobas2020, Maskara2021} and references therein. In what follows we study the step-like stroboscopic Floquet protocol for the kicked $3$-state Potts model discussed in Section~\ref{sec:Hamiltonian}.

\subsection{Kicked $3$-state Potts model}

We now consider the time-dependent Hamiltonian of the $3$-state Potts model subject to periodically kicked transverse field~$H_1$:
\be \label{H_kicked_Potts}
	H(t) =  -J H_1 - f T \sum_{m \in {\mathbb Z}} \delta(t - m T)  H_2,
\ee
where $H_{1,2}$ are given by Eq.~(\ref{Potts_H1_H2_J}). Thus, the time-dependent Hamiltonian~$H(t)$ in Eq.~(\ref{H_kicked_Potts}) is of the form~(\ref{H_kicked_general}), with~$g_1 = - J$ and~$g_2 = - f$.
Then, the one-period stroboscopic Floquet operator is given by Eq.~(\ref{step-like_protocol_general}) and reads as
\be \label{Floquet_unitary_general}
	U_F(T) =  e^{i f T H_2} e^{i J T H_1}.
\ee
It corresponds to the evolution for time~$T$ with the interaction Hamiltonian~$-J H_1$, followed by an instantaneous kick by the uniform transverse field~$-f H_2$. We are interested in the state of the system 
\be
	\ket{\psi(kT)} = U_F^k(T) \ket{\psi(0)}
\ee
after~$k$ periods of the protocol. 

Let us denote by~$\tilde H_1$ the interaction part of the Hamiltonian with the central link (i.e. that between the sites $N$ and $N+1$) being switched off~\cite{note_unequal_split}:
\be \label{tilde_H1}
	\tilde H_1 = H_1 - \left( X_N^{\dag} X_{N+1} + X_N X_{N+1}^{\dag} \right).
\ee
Then, we rewrite the one-period evolution operator by separating the part~$V_0$ that acts only in the middle of the chain:
\be \label{Floquet_unitary_tilde_U_V0}
	U_F(T) =  e^{i f T H_2} e^{i J T \tilde H_1 } V_0(T) \equiv \tilde U(T) V_0(T),
\ee
where we introduced unitary operators~$V_0$ and $\tilde U$. The former is given by
\be \label{V_0_def}
	V_0  = \exp \left\{ i J T \, \left( X_N^{\dag} X_{N+1} + X_N X_{N+1}^{\dag} \right) \right\}
\ee
and acts non-trivially only at the $N$-th and $(N+1)$-th sites, i.e. over the central link of the chain. On the contrary, the operator
\be \label{tilde_U_tilde_H}
	\tilde U(T) = e^{i f T H_2} e^{i J T \tilde H_1}
\ee
acts non-trivially on the left and right halves of the chain and {\it not} across the central link. 
Therefore,~$\tilde U$~does not entangle the left and right halves with each other and can be written in the factorized form
\be \label{tilde_U_LR}
	\tilde U(T) = \tilde U_L(T) \tilde U_R(T),
\ee
where $\tilde U_{L(R)}$ acts on the left (right) half of the lattice and~$[\tilde U_L, \tilde U_R] = 0$. Then, we rewrite the one-period Floquet operator in~Eq.~(\ref{Floquet_unitary_tilde_U_V0}) as
\be \label{U_decomposition}
	U_F = \tilde U V_0 = V_1 \tilde U, \qquad V_1 = \tilde U V_0 \tilde U^{\dag},
\ee
were for the sake of readability we omitted the $T$-dependence. Similarly, for two periods we have $U_F^2 = \tilde U V_0 V_1 \tilde U$, which can be written as
\be \label{V_2}
	U_F^2 = V_1 V_2 \tilde U^2, \qquad V_2 = \tilde U V_1 \tilde U^{\dag}.
\ee
One can easily check that the evolution operator for $k$~periods becomes
\be \label{U_k_general}
	U_F^k =  V_1 \ldots V_{k}  \, \tilde U^k, \qquad V_l = \tilde U V_{l-1} \tilde U^{\dag}.
\ee 
The unitary operators $V_k$ act non-trivially on both halves of the chain and entangle them. However, for some specific values of $f T$ and $J T$ the resulting entanglement has a very simple structure, as we show below.

\subsection{Explicit form of $V_k$} \label{S:V_k_explicit}

Let us now find the explicit form of the operators $V_k$. 
We first rewrite the operator $V_1$ from Eq.~(\ref{U_decomposition}) as
\be \label{V_1}
\begin{aligned}
	V_1 &= \tilde U V_0 \tilde U^{\dag} = e^{i f T \text{ad}_{H_2}} V_0\\
	&=\prod_{j=N}^{N+1}\prod_{n=1}^{2} \sum_{k=0}^{\infty} \frac{(i f T)^k}{k!} \text{ad}_{ {\cal J}_j^{(0,n)} }^k \; V_0
\end{aligned}
\ee
where $\text{ad}_X^k Y \equiv \left[ X, \bigl[X , \ldots [X,Y] \bigr] \right]$ is a $k-$fold nested commutator. 
In writing Eq.~(\ref{V_1}) we took into account that $[\tilde H_1, V_0] = 0$ and all terms in~$H_2$ commute with each other, since the operators~${\cal J}_j^{\boldsymbol{m}}$ commute on different sites. We also used a well known identity $e^{X} Y e^{-X} = e^{\text{ad}_X} Y$, valid for any $X$ and $Y$ in a Lie algebra [${\mathfrak sl}(3, {\mathbb C})$ in our case]. To simplify Eq.~(\ref{V_1}) it is convenient to expand the exponential in $V_0$. From Eq.~(\ref{V_0_def}) we have (for details, see Appendix~\ref{A:V0_exp_series})
\be \label{V_0_exp_series}
	V_0 = \mu {\mathds 1} + \nu \left( {\cal J}_{N}^{(1,0)}{\cal J}_{N+1}^{(2,0)} + {\cal J}_{N}^{(2,0)}{\cal J}_{N+1}^{(1,0)} \right),
\ee
were we took into account Eq.~(\ref{Z_X_to_J}) for the definition of the operators~${\cal J}_{j}^{{\bs m}}$. The coefficients in Eq.~(\ref{V_0_exp_series}) are given by
\be  \label{mu_nu}
	\mu = \nu + e^{- i T J}, \qquad \nu = e^{- i T J} (e^{3i T J}-1)/3.
\ee
It is now straightforward to calculate~$V_1$. For generic values of~$f T$ the calculation of the adjoint action in Eq.~(\ref{V_1}) can be found in~Appendix~\ref{A:V1_V2_Vk_derivation}. Importantly, 
the expression for~$V_1$ becomes especially simple for~$f T = \alpha_m$, where we denoted
\be \label{alpha_m_special_value}
	\alpha_m = \frac{2\pi}{9}(3l - m),
\ee
with $l \in {\mathbb Z}$ and $m \in \{ 0,1,2 \}$. 
In this case, $V_1$ reads as
\be \label{V_1_res_J_m}
	V_1  = \mu {\mathds 1} + \nu  \left({ \cal J}_N^{(1,m)}{\cal J}_{N+1}^{(2,2m)} + {\cal J}_N^{(2,2m)}{\cal J}_{N+1}^{(1,m)} \right).
\ee
Obviously, the case $m=0$ is trivial, since it results in~$V_1 = V_0$. 

Moreover, one can show that the operator~$V_2$ from Eq.~(\ref{V_2}) also acquires a compact form for 
\be \label{T_1=T_2}
	f T = J T = \alpha_m,
\ee
where~$\alpha_m$ is given by Eq.~(\ref{alpha_m_special_value}). Using the results of Appendices~\ref{A:adjoints} and~\ref{A:V1_V2_Vk_derivation}, we find
\be \label{V_2_res_J}
\begin{aligned}
	V_2 = \mu {\mathds 1} + \nu \Bigl(& {\cal J}_{N-1}^{(1, m)} {\cal J}_{N}^{(0, m)} {\cal J}_{N + 1}^{(0, 2 m)} {\cal J}_{N + 2}^{(2, 2 m)} \\
	&+ {\cal J}_{N-1}^{(2, 2m)} {\cal J}_{N}^{(0, 2m)} {\cal J}_{N + 1}^{(0, m)} {\cal J}_{N + 2}^{(1, m)} \Bigr).
\end{aligned}
\ee
Similarly,~under the conditions $J T = f T = \alpha_m$ 
and $2 \leq k \leq N$ one obtains the following expression for $V_k$: 
\be \label{V_k_res_J}
\begin{aligned}
	V_k = &\mu {\mathds 1}  + \nu  \Bigl( {\cal J}_{N-k+1}^{(1,m)} \, {\cal J}_{N-k+2}^{(0,m)} \ldots {\cal J}_{N}^{(0,m)} \\
	&\quad \times {\cal J}_{N+1}^{(0,2m)} \ldots {\cal J}_{N+k-1}^{(0,2m)} \, {\cal J}_{N+k}^{(2,2m)} +\text{H.c} \Bigr),
\end{aligned}
\ee
where in the conjugated term (denoted by ``H.c.''), one simply makes a replacement $m \leftrightarrow 2m$ in the upper indices of ${\cal J}_{j}^{(p,q)}$.
In terms of the operators~$X_j$ and~$Z_j$ one can write~$V_k$ as
\be \label{V_k_res}
\begin{aligned}
	V_k =&  \mu {\mathds 1}  + \nu  \omega^{2m} \Bigl( {X}_{N-k+1} {Z}_{N-k+1}^{m} \ldots {Z}_{N}^{m}  \\
	&  \,\times Z_{N+1}^{2m} \ldots Z_{N+k-1}^{2m} X_{N+k}^2 \, Z_{N+k}^{2m}  + \text{H.c.}\Bigr),
\end{aligned}
\ee 
which follows from Eq.~(\ref{Z_X_to_J}).
We thus see that for~$J T = f T = 2\pi  (3l - m) /9$ the form of the operator~$V_k$ from Eq.~(\ref{V_k_res}) is quite simple, as well as its action on the chain. It only changes the internal state of qutrits on the sites~$N-k+1$ and~$N+k$, whereas on the rest of the chain~$V_k$ either produces an extra phase factor or acts trivially. 

Interestingly, for $k > N$ the form of $V_k$ exhibits a peculiar structure. 
Before we proceed, let us rewrite~$V_k$ for later convenience as
\be \label{V_k_three_terms}
	V_k = \mu {\mathds 1} + \nu \left( {\cal V}_k + {\cal V}_k^{\dag} \right), 
\ee
which can be always done since~$V_0$ has this form. Moreover, the operator ${\cal V}_k$ in Eq.~(\ref{V_k_three_terms}) can be always written as 
\be
\begin{aligned}
	 {\cal V}_k \equiv {\cal W}_k \tilde {\cal W}_k =&  \prod_{j=1}^N {\cal J}_j^{(p_j(k),q_j(k))} \\
	& \times  \prod_{j=N+1}^{2N} {\cal J}_{j}^{(2p_{2N+1-j}(k), 2q_{2N+1-j}(k))},
\end{aligned}
\ee
because the upper indices of ${\cal J}_{j}^{(p_j(k),q_j(k))}$ in ${\cal V}_k$ are symmetric with respect to reflection across the $N$th link of the chain. For brevity, let us focus on the structure of~${\cal W}_k = \prod_{j=1}^N {\cal J}_j^{(p_j(k),q_j(k))}$.
Using the results of Appendices~\ref{A:adjoints} and~\ref{A:V1_V2_Vk_derivation}, we obtain the following expressions for~${\cal W}_k$ with $k>N$:
\be \label{W_k_explicit}
\begin{aligned}
	{\cal W}_{N + l} & = \prod_{j=1}^{l-1}{\cal J}_j^{(0,2m)} \; {\cal J}_l^{(1,2m)} \; \prod_{j=l+1}^N{\cal J}_{j}^{(0,m)}, \\
	{\cal W}_{2N + l} & = \prod_{j=1}^{N-l}{\cal J}_j^{(0,2m)} \; {\cal J}_{N-l+1}^{(1,0)} \; \prod_{j=N-l+2}^N{\mathds 1}_j, \\
	{\cal W}_{3N + l} & = \prod_{j=1}^{l-1}{\cal J}_j^{(0,m)} \; {\cal J}_{l}^{(1,m)} \; \prod_{j=l+1}^N{\mathds 1}_j, \\
	{\cal W}_{4N + l} & = \prod_{j=1}^{N-l}{\cal J}_j^{(0,m)} \; {\cal J}_{N-l+1}^{(1,2m)} \; \prod_{j=N-l+2}^N{\cal J}^{(0,2m)}, \\
	{\cal W}_{5N + l} & = \prod_{j=1}^{l-1}{\mathds 1}_j \; {\cal J}_{l}^{(1,0)} \; \prod_{j=l+1}^N{\cal J}^{(0,2m)}, \\
	{\cal W}_{6N + l} & = \prod_{j=1}^{N-l}{\mathds 1}_j \; {\cal J}_{N-l+1}^{(1,m)} \; \prod_{j=N-l+2}^N{\cal J}^{(0,m)}, \\
\end{aligned}
\ee
where~$1\leq l \leq N$. The structure of~${\cal W}_k$ in Eq.~(\ref{W_k_explicit}) is fairly complicated and one can see that the upper indices of ${\cal J}_{j}^{(p_j(k),q_j(k))}$ change completely across~${\cal W}_k$ several times as~$k$ increases from~$N$ to $6N$ [recall that $N$ is {\it half} the length of the chain]. Quite remarkably, the expression for~${\cal W}_{6N+l}$ from Eq.~(\ref{W_k_explicit}) coincides with that for~$W_l$ with $1\leq l \leq N$, as can be seen from Eq.~(\ref{V_k_res_J}). This means that the operators~$V_k$ are periodic with respect to~$k$ and one has
\be \label{V_k_peroiodic}
	V_{6N + k} = V_{k}.
\ee
Then, taking into account Eq.~(\ref{U_k_general}) we have $V_{6N + k} = \tilde U^{6N} V_k \tilde U^{\dag \; 6N}$, and one concludes that~$[\tilde U^{6N}, V_k]=0$ for any $k$. Our detailed analysis shows that in fact one has 
\be
	\tilde U^{6N} = {\mathds 1},
\ee
meaning that under the condition~(\ref{T_1=T_2}) the unitary operator~$\tilde U$ from Eq.~(\ref{tilde_U_tilde_H}) is a permutation. This has far reaching consequences as one should be able to find explicitely the spectrum of the total Floquet operator~$U_F$ from Eq.~(\ref{Floquet_unitary_general}). However, this lies beyond the scope of the present work and we leave it to future studies.

In the following we show explicitly that in the regime $f T = J T = \alpha_m$, with $\alpha_m$ given by Eq.~(\ref{alpha_m_special_value}), the Floquet protocol described in this Section leads to the generation of long-range entanglement between the pairs of qutrits, and the resulting entanglement has a very simple form.

\subsection{Long-range entangled state generation} \label{S:entanglement_generation}

The protocol consists of the preliminary state preparation and the generation of long-range entanglement itself. In the first stage, we start from an initial polarized state~$\ket{\psi_0}$, i.e. the product state in which all qutrits are in one and the same internal state~\cite{note_initial_state}. For instance, let us assume that every qutrit is initialized in the state~$\ket{0}$. Thus, the initial state for the protocol reads
\be \label{polarized_init_state}
	\ket{\psi(0)} = \otimes_{j=1}^{2N}\,\ket{0}_j.
\ee
Then, the state preparation procedure consists of evolving the state~$\ket{\psi(0)}$ for~$k$ periods by the two-step Floquet protocol with the one period Floquet unitary~$\tilde U^{\dag}(T)$:
\be
	 | \tilde \psi (k T) \rangle = \tilde U^{\dag \, k}(T) \ket{\psi(0)} 
	  = |\tilde \Psi(kT)\rangle_L \otimes |\tilde \Psi(kT)\rangle_R, 
\ee
where~$\tilde U(T)$ is given by Eq.~(\ref{tilde_U_tilde_H}) and we took into account that it factorizes according to Eq.~(\ref{tilde_U_LR}).  From~Eq.~(\ref{tilde_U_tilde_H}) we have
\be \label{tilde_U_dag}
	\tilde U^{\dag} (T)= e^{-i J T \tilde H_1} e^{- i f T H_2},
\ee
which corresponds to the stroboscopic evolution with the Hamiltonian
\be
	H^{\prime}(t) = f H_2 + J \sum_{n \in {\mathbb Z}} T \delta(t - n T) \tilde H_1,
\ee
where~$\tilde H_2$ is given by Eq.~(\ref{tilde_H1}) and~$H_2$ by Eq.~(\ref{Potts_H1_H2_XZ}). In other words, each period of the state preparation stage consists of evolving the state for time~$T$ with the Hamiltonian~$f H_2$ (corresponding to the transverse field), followed by an instantaneous kick with the Hamiltonian~$J \tilde H_1$, which corresponds to the interaction between the nearest neighbour spins on all links except for the central one~\cite{note_unequal_split}.

At the next step, we perform the Floquet protocol with the full evolution operator~$U_F(T)$ from Eqs.~(\ref{Floquet_unitary_general}) and~(\ref{Floquet_unitary_tilde_U_V0}). 
After the first period of the protocol we obtain
\be
\begin{aligned}
	\ket{\psi(T)} &= U_F(T) \,  | \tilde \psi (k T) \rangle 
	= V_1  \bigl[ \tilde U^{\dag}(T) \bigr]^{k-1} \ket{\psi(0)}\\
	&= V_1 |\tilde \Psi((k-1)T)\rangle_L \otimes |\tilde \Psi((k-1)T)\rangle_R.
\end{aligned}
\ee
Then, after $k$ periods we arrive at the state
\be \label{psi_k_gen}
	\ket{\psi(kT)} = U_F^k(T) \, | \tilde \psi (k T) \rangle =  V_1 \ldots V_{k} \ket{\psi(0)},
\ee
where for~$U_F^k$ we used Eq.~(\ref{U_k_general}). We thus see that the evolution generated by the operator~$\tilde U^{\dag\,k}$ is eliminated, and the resulting state~$\ket{\psi(kT)}$ in Eq.~(\ref{psi_k_gen}) is fully determined by the action of the string~$V_1 \ldots V_k$ on the initial state~$\ket{\psi(0)}$. Note that Eq.~(\ref{psi_k_gen}) is completely general and is valid for arbitrary values of~$f T$ and~$J T$.

\begin{widetext}

\begin{figure}
	\includegraphics[width=\linewidth]{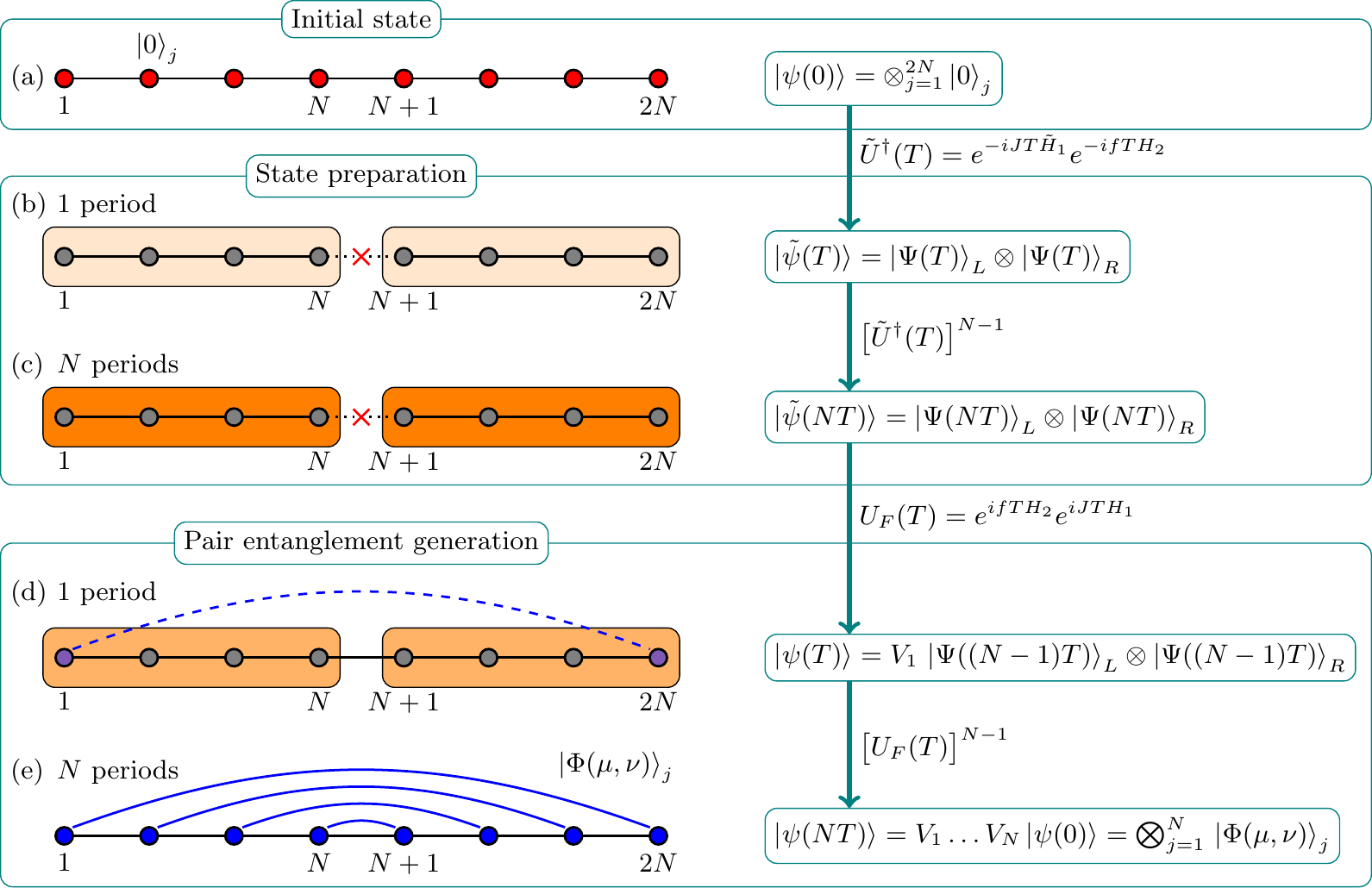}
\caption{Schematic illustration of the protocol for generating long-range entanglement in the $3$-state Potts model. 
The protocol starts with the initial state (a) and consists of the preliminary state preparation scheme [shown in (b) and (c)] followed by the procedure for generating non-local entanglement [displayed in (d) and (e)] between distant qutrit pairs.
(a) At $t=0$ every qutrit is in one and the same state~$\ket{0}_j$, so that the chain initially is in the polararized product state~$\ket{\psi(0)}$. At the first iteration of the state preparation scheme (b), the central link of the chain is switched off (shown by the red cross on the link) and the two halves of the chain are evolved independently with the unitary operator~$\tilde U^{\dag}(T)$, which is given by Eq.~(\ref{tilde_U_dag}). This produces a bipartite state~$| \tilde \psi(T) \rangle$ shown by two light colored blocks in (b). Then, after $N-1$ more iterations of the state preparation scheme (c),  the chain is in the bipartite state $| \tilde \psi(NT) \rangle$. The two halves of the chain are still factorized, but each of them is now highly entangled, as illustrated by the darker colored  blocks in (c). This completes the state preparation scheme and one proceeds with the protocol generating entanglement between qutrit pairs. At the first period the protocol (d), one first reduces the amount of entanglement {\it inside} the left and right halves of the chain by the operator~$U_F(T)$ and produces the state~$| \tilde \psi((N-1)T) \rangle$ [note that it is represented by the blocks of lighter color as compared to (c)]. In addition, by the operator~$V_1$ one creates a two-particle entanglement {\it between} the left and right halves of the chain [blue dashed arc in (d)].
The result of every next period is also two-fold: one further disentangles separately the states of  the left and right halves of the chain and creates another two-site entanglement between them. As a result, after~$N$ periods of the protocol (e) the final state~$\ket{\psi(NT)}$ is simply the product of maximally entangled two-qutrit states~$\ket{\Phi(\mu,\nu)}_j$ [blue solid arcs in (e)].
}
\label{F:Potts_entanglement_gen}
\end{figure}

\end{widetext}

In order to proceed with constructing the explicit form of the state~$\ket{\psi(kT)}$ from Eq.~(\ref{psi_k_gen}), we choose~$J T = f T = 2\pi (3l-m)/9$, so that the operators~$V_k$ have the simple form given in Eq.~(\ref{V_k_res}). In this case, one can clearly see that by applying $V_{k}$ to the initial state~$\ket{\psi(0)}$ we entangle only the sites $N-k+1$ and $N+k$:
\be \label{max_ent_state}
	V_{k} \ket{\psi(0)} = \ket{ \Phi(\mu, \nu)}_{k} \bigotimes_{j=1}^{2N}{}^{\prime}\,\ket{0}_j,
\ee
where  the prime means that the tensor product does not include sites~$j = N - k + 1$ and $j = N+k$. In Eq.~(\ref{max_ent_state}) we denoted by $\ket{\Phi(\mu, \nu)}_{j}$ the maximally entangled two-qutrit Bell-like state
\be \label{Phi_explicit_form}
\begin{aligned}
	&\ket{\Phi (\mu, \nu)}_{j} = \mu \ket{0}_{N-j+1} \ket{0}_{N+ j} \\
	&+ \omega^{2m} \nu \left( \ket{1}_{N-j+1} \ket{2}_{N+ j}
	 + \ket{2}_{N-j+1} \ket{1}_{N+ j} \right),
\end{aligned}
\ee
with~$m=1,2$ and $\mu$ and $\nu$ given by Eq.~(\ref{mu_nu}). One can easily see that the state~$\ket{\Phi(\mu, \nu)}_{j}$ has the Schmidt rank~$r=3$. Taking into account that for~$J T = 2\pi  (3l - m) /9$ we have~$|\mu| = |\nu| = 1/\sqrt{3}$, it is also easy to check that the partial trace of~$\ket{\Phi(\mu, \nu)}_{j}\bra{\Phi(\mu, \nu)}$ with respect to either of the two subspaces gives~${\mathds 1}/3$. Therefore, the state~$\ket{\Phi(\mu, \nu)}_{j}$ in Eq.~(\ref{Phi_explicit_form}) is indeed a maximally entangled one~\cite{Sych2009}.
Then, from~Eqs.~(\ref{psi_k_gen}) and (\ref{max_ent_state}) we immediately obtain the final state of the chain after~$k$ periods of the Floquet protocol:
\be \label{final_state}
	\ket{\psi(kT)} = \bigotimes_{j = 1}^{N-k}\,\ket{0}_j \bigotimes_{j=1}^{k}\,\ket{\Phi (\mu, \nu)}_j \bigotimes_{j = N+k+1}^{2N}\,\ket{0}_j ,
\ee
which is build of non-local qutrit pairs with a long-range entanglement. The result in~Eq.~(\ref{final_state}) is valid for the number of periods~$k \leq N$, otherwise there are obviously not enough qutrit pairs to entangle. In particular, for~$k = N$ the final state consists of $N$ maximally entangled qutrit pairs, which are symmetrically distributed around the middle of the chain. We schematically illustrate the Floquet protocol in Fig.~\ref{F:Potts_entanglement_gen}.
Let us emphasize that the suggested Floquet protocol works properly only under a specific choice of the system parameters. Namely, the state~(\ref{final_state}) is obtained under the condition $fT = JT = \alpha_m$, with $\alpha_m$ given in Eq.~(\ref{alpha_m_special_value}),
which guarantees the remarkably simple structure of the operators~$V_k$ in Eq.~(\ref{V_k_res}) and, consequently, that of the final state~$\ket{\psi(kT)}$ in Eqs.~(\ref{psi_k_gen}) and~(\ref{final_state}). 

\begin{widetext}

\begin{figure}[t]
\subfloat{
	\includegraphics[width=0.45\linewidth]{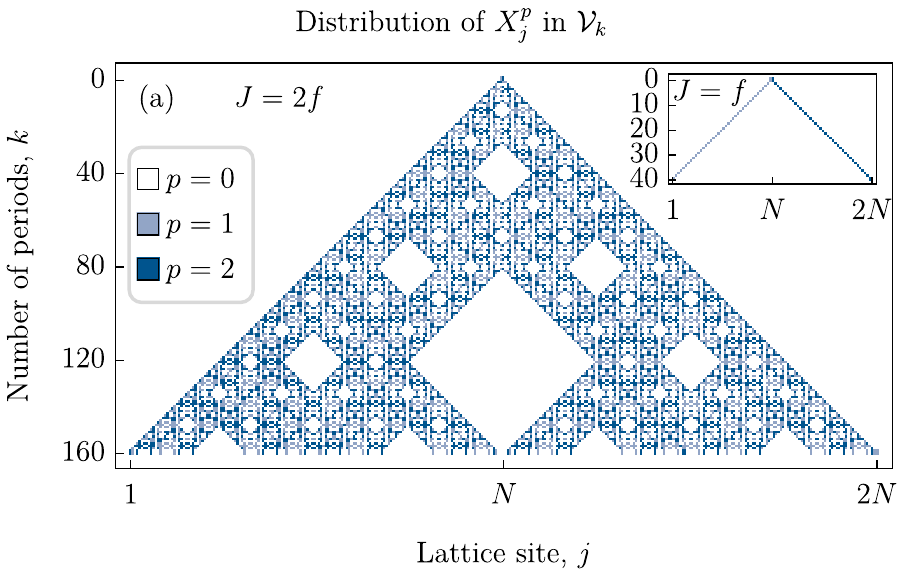}
	}
	\hfill
\subfloat{
	\includegraphics[width=0.45\linewidth]{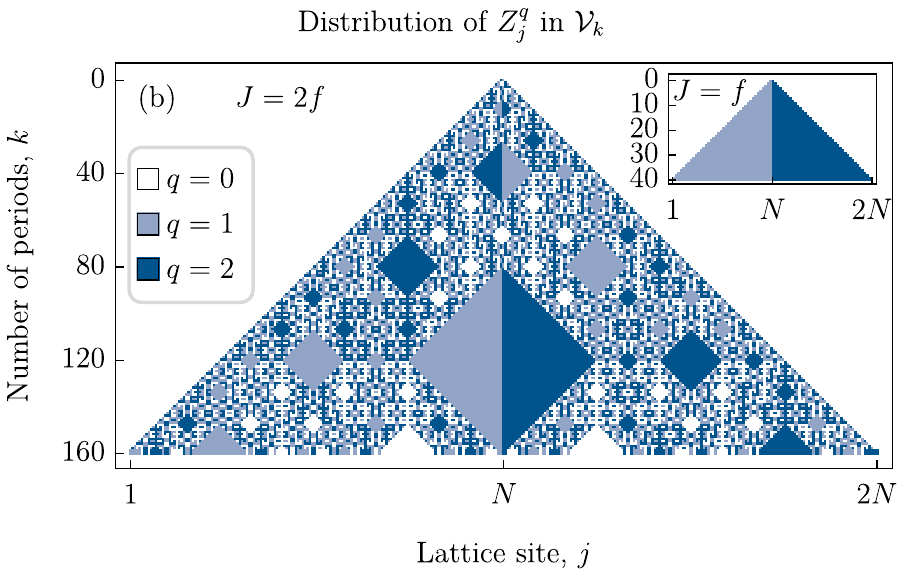}
	}
	\caption{Structure of~${\cal V}_k$ in Eqs.~(\ref{V_k_three_terms}), (\ref{calV_k_powers}). {\it Panel}~(a): Distribution of $X_j^p$ with $1\leq j \leq 2N$ in~${\cal V}_k$ for~$fT=2\pi/9$ and~$JT=4\pi/9$ (main panel), chain length of~$2N=320$ sites, and the number of periods $0\leq k \leq N=160$. 
	The color of a pixel with coordinates~$(j,k)$ encodes the power~$p$ of the operator~$X_j^p$ in the expression for~${\cal V}_k$ from Eq.~(\ref{calV_k_powers}). 
	The inset shows the same distribution for $fT=JT=2\pi/9$ and $2N= 80$. 
	{\it Panel}~(b): Distribution of $Z_j^q$ with $1\leq j \leq 2N$ in~${\cal V}_k$. The color of a~$(j,k)$-th pixel indicates the power~$q$ of~$Z_j^q$~in the expression for~${\cal V}_k$. The parameters on the main panel and the inset are the same as the corresponding ones on the panel~(a). Note that if one instead chooses~$fT=2\pi/9$ and~$JT=4\pi/9$ for both main panels, the distribution of~$X_j^p$ remains the same, whereas in the distribution of~$Z_j^q$ one simply swaps~$q = 1$ and $q = 2$. Likewise, taking for the insets~$f T = JT = 4\pi/9$, the inset in panel~(a) does not change, whereas the inset in panel~(b) gets reflected.
	 } \label{F:fractal}
\end{figure}

\end{widetext}

Before we complete the discussion of the long-range entanglement generation, we also would like to mention a somewhat unrelated but peculiar observation. Imagine that instead of considering the regime in~Eq.~(\ref{T_1=T_2}) one takes, say, $fT = 2\pi/9$ and $J T = 4 \pi/9$. In this case, the operator~$V_k$ still contains only three terms, just like in Eqs.~(\ref{V_k_res}) and~(\ref{V_k_three_terms}). 
For $f T = J T = \alpha_m$~one can immediately read off the form of ${\cal V}_k$ from Eq.~(\ref{V_k_res}), and it is extremely simple. On the contrary, for $fT = \alpha_{2m}$  and $J T = \alpha_{m}$ (or vice versa), the structure of~${\cal V}_k$ in Eq.~(\ref{V_k_three_terms}) turns out to be quite complex and it drastically changes with~$k$, as we discuss in more detail in Appendix~\ref{A:V1_V2_Vk_derivation} (see Eq.~\ref{V_4_5_6_f=2J}). Neglecting the phase factor, one can write~${\cal V}_k$ as
\be \label{calV_k_powers}
	{\cal V}_k \sim \prod_{j=k-1}^0 {X}_{N-j}^{p_{j+1}} {Z}_{N-j}^{q_{j+1}} \prod_{j=1}^{k} {X}_{N+j}^{2 p_{j}} {Z}_{N+j}^{2 q_j},
\ee
so that the operator content of~$V_k$ in Eq.~(\ref{V_k_three_terms}) is characterized by two $2N$-dimensional vectors~${\boldsymbol P}_k$ and ${\boldsymbol Q}_k$, containing the powers of $X_j$ and $Z_j$, correspondingly, on all lattice sites:
\be \label{vector_X_powers}
	{\boldsymbol P}_k = \{p_{1}(k), \ldots, p_N(k),  2 p_N(k), \ldots,  2 p_1(k)\},
\ee
and similarly for~${\boldsymbol Q}_k$. In the course of the Floquet protocol, the components of ${\boldsymbol P}_k$ and ${\boldsymbol Q}_k$ are updated via Eq.~(\ref{J_indices_lin_transform}), which follows from the relation~$V_k = \tilde U V_{k-1} \tilde U^{\dag}$ in Eq.~(\ref{U_k_general}). Then, in order to gain insight on the structure of~$V_k$ for $fT = 2\pi/9$  and $J T = 4 \pi/9$, let us plot the components of~${\boldsymbol P}_k$ and~${\boldsymbol Q}_k$ for different values of~$k$ and see how they change across the chain as the number of periods~$k$ increases. 
Our findings are illustrated in Fig.~\ref{F:fractal}. One clearly sees that the components of~${\boldsymbol P}_k$ and~${\boldsymbol Q}_k$ form a fractal pattern, which exhibits a large-scale structure resembling the Sierpi\'nski carpet. It would be interesting to investigate the origin of this fractal behaviour, as well as its possible physical consequences on the Floquet dynamics. However, this lies beyond the scope of the present paper and we leave it to future studies. 

Returning to the regime of equal $fT$ and $JT$, one may ask what is the physical reason behind the fine-tuning requirement in Eq.~(\ref{T_1=T_2}), which has to be satisfied for the Floquet protocol to generate the state with long-range entangled pairwise entanglement between qutrits. In Section~\ref{sec:Floquet_integrability} we argue that Eq.~(\ref{T_1=T_2}) is nothing else than the {\it integrability condition} for the stroboscopic Floquet protocol~(\ref{Floquet_unitary_general}). In this view it is quite natural that the protocol creates a state of a simple form, instead of simply heating the system up to infinite temperature.

\section{Integrability of the Floquet protocol} \label{sec:Floquet_integrability} \label{S:integrability}

\subsection{General remarks and relation to the Temperley-Lieb algebra}

We now show that the protocol consider in Section~\ref{sec:Floquet_protocol} can be viewed as a special case of a more general Floquet protocol. 
The reason is that the $3$-state Potts model is a {\it representation} of a Hamiltonian that belongs to the so-called Temperley-Lieb algebra. In this Section we briefly overview the Temperley-Lieb algebra, construct a stroboscopic two-step Floquet protocol using the generators of the algebra, and discuss the Floquet integrability of the protocol.

Let $u_j$ with $j=1,\ldots,L-1$ be the generators of the Temperley-Lieb algebra $TL_{L}(\beta)$, where $\beta$ is a complex parameter. 
The generators satisfy the defining relations
\begin{subequations} \label{TL_rels}
\begin{align}
	&u_j^2 = \beta u_j, \label{TL_rels_square} \\
	&u_j u_{j\pm1} u_j = u_j, \label{TL_rels_brainding} \\
	&[ u_i,  u_j] = 0, \quad |i-j|>1 \label{TL_rels_commutativity}.
\end{align}
\end{subequations}
The elements (also called {\it words}) of $TL_{L}(\beta)$ are obtained by multiplying the generators~$u_j$ in all possible ways. A word~$w$ is called {\it reduced} if it cannot be shortened with the help of the relations~(\ref{TL_rels}). Every reduced word $w \in TL_{L}(\beta)$ can be written in the {\it Jones normal form}~\cite{Ridout2014}, namely as a sequence of decreasing sequences of the generators:
\be \label{TL_reduced_word}
	w =  \left( u_{j_1} u_{j_1-1} \ldots u_{k_1} \right)\ldots \left( u_{j_r} u_{j_r-1} \ldots u_{k_r} \right),
\ee
where $0 < j_1 < \ldots < j_r < L$ and $0 < k_1 < \ldots < k_r < L$. It can be shown that the generator with the largest index appears in~$w$ only once. All reduced words formed of the generators~$\{u_j\}_{j=1}^{L-1}$ span the basis in~$TL_L(\beta)$. Thus, the Temperley-Lieb algebra is finite dimensional and one can show that its dimensionality is 
\be \label{TL_dim}
	\text{dim }TL_L(\beta) = \frac{1}{L+1}\binom{2 L}{L} = C_{L},
\ee
which is the $L$th Catalan number. Let us note that the defining relations~(\ref{TL_rels}) can be equivalently formulated using the so-called Gr\"obner-Shirshov basis~\cite{TL_Groebner_basis}. To do so, one introduces the operators~$u_{i,j}$ defined as
\be \label{TL_Groebner_def}
\begin{aligned}
	&u_{i,i} = u_i, \qquad u_{i, i+1} = 1, \\
	&u_{i,j} = u_{i} u_{i-1} \ldots u_{j}, \;\; i \geq j.
\end{aligned}
\ee
Then, one replaces the relation~(\ref{TL_rels_brainding}) with the following two:
\be \label{TL_Groebner_rels}
\begin{aligned}
	&u_{i,j} \, u_i = u_{i-2, j} \, u_i, \\
	&u_j \, u_{i,j} = u_j \, u_{i, j+2} , \\
\end{aligned}
\ee
where~$i>j$. One can easily check that Eqs.~(\ref{TL_rels_square}), (\ref{TL_rels_commutativity}), (\ref{TL_Groebner_def}), and (\ref{TL_Groebner_rels}) are equivalent to the standard form of the defining relations in~Eq.~(\ref{TL_rels}). However, the former are often much more convenient in practice. 

Quite remarkably, the Temperley-Lieb algebra possesses numerous representations that correspond to various paradigmatic physical models, such as the TFIM, spin-$1/2$ XXZ spin chain, and the $n$-state Potts model~\cite{Nichols2006}.  
We are interested in the representation corresponding to the 3-state Potts model on a chain with $M$ sites and open boundary conditions:
\be \label{TL_Potts_rep}
\begin{aligned}
	&u_{2j} = \dfrac{1}{\sqrt{3}}\left(1+ X_j X_{j+1}^{\dagger} + X_j^{\dagger} X_{j+1}\right), \;\; 1 \leq j < M,\\
	&u_{2j -1} = \dfrac{1}{\sqrt{3}}\left(1 + Z_j + Z_j^{\dagger}\right), \qquad 1 \leq j \leq M.
\end{aligned}
\ee
One can easily check that the operators~$u_j$ in Eq.~(\ref{TL_Potts_rep}) satisfy the defining relations in Eq.~(\ref{TL_rels}) with $\beta = \sqrt{3}$ and thus form a representation of $TL_{2M}(\sqrt{3})$. 

We now consider the following linear combinations belonging to $TL_{2M}(\beta)$:
\be \label{TL_H1_H2}
	H_1 = \sum_{j=1}^{M-1} u_{2j}, \qquad H_2 = \sum_{j=1}^{M} u_{2j-1}.
\ee
Assuming that~$u_{j}$ are Hermitian, we treat $H_{1,2}$ as abstract Hamiltonians for which one can use any Hermitian representation, in particular the one in Eq.~(\ref{TL_Potts_rep}). We then construct a stroboscopic two-step Floquet protocol
\be \label{TL_U_F}
	U_F = e^{-i T_2 H_2} e^{-i T_1 H_1},
\ee
where $T_k$ is either the time period over which the dynamics is governed by $H_k$, as in Eq.~(\ref{step-like_protocol_general_2}), or~$T_k \equiv g_k T$ as in Eq.~(\ref{step-like_protocol_general}) corresponding to the kicked protocol. In the latter case~$T_k$ are allowed to be negative. Thus, taking~$M=2N$, $T_1 = - \sqrt{3} J T$, $T_2 = - \sqrt{3} f T$, and using the representation~(\ref{TL_Potts_rep}), we reduce Eq.~(\ref{TL_U_F}) to the Floquet protocol in Eq.~(\ref{Floquet_unitary_general}) up to a constant phase.  However, in what follows we mostly work with the general case in Eq.~(\ref{TL_U_F}), hereinafter referred to as the Temperley-Lieb algebraic Floquet protocol. 

\subsection{Integrability of the Temperley-Lieb algebraic Floquet protocol and its conservation laws}

Note that the Floquet operator in Eq.~(\ref{TL_U_F}) can be written in the form
\be \label{U_F_T_matrix}
	U_F = \prod_{j=1}^{M}\left( {\mathds 1} + x_2 u_{2j-1} \right) \prod_{j=1}^{M-1}\left( {\mathds 1} + x_1 u_{2j} \right),
\ee
where we used Eq.~(\ref{TL_rels}) and denoted~$x_k = (e^{- i \beta T_k } - 1)/\beta$. In this form, the Floquet evolution operator~$U_F$ resembles the transfer matrix of a two-dimensional classical {\it integrable} lattice model. This similarity suggests that the Floquet protocol~(\ref{U_F_T_matrix}) is also integrable in some sense~\cite{Gritsev2017}. This is indeed the case, and the notion of integrability in the context of Floquet dynamics should be understood in the following way.
Let us rewrite Eq.~(\ref{TL_U_F}) in the form of a single exponential [as in Eq.~(\ref{stroboscopic_Floquet_unitary_single_exp})]:
\be \label{TL_U_F_H_F}
	U_F = e^{ - i T H_F },
\ee
where $H_F$ is an effective Floquet Hamiltonian. Then, for an integrable stroboscopic Floquet protocol with the one-period evolution operator~$U_F$ there is a macroscopically large number of conserved quantities (charges)~$Q_n$ that commute with the effective Hamiltonian~$H_F$ and with each other:
\be
	[Q_n,  H_F] = 0, \qquad [Q_n, Q_m] = 0.
\ee
Obviously, due to Eq.~(\ref{TL_U_F_H_F}),~the charges $Q_n$ also commute with the Floquet evolution operator~$U_F$, which eliminates the need to calculate~$H_F$ explicitely.
Importantly, the conserved charges are required to be local, i.e. expressible as a linear combination of terms with a finite support:
\be \label{TL_Qn_ansatz_general}
	Q_n = \sum_{ l \lesssim n} \sum_j q_{j, j+1, \ldots j+l}^{(n)}.
\ee
The support of~$q_{j, j+1, \ldots j+l}^{(n)}$ increases with~$n$, but for a local charge $Q_n$ remains finite. For the Temperley-Lieb algebraic Floquet protocol the operators~$q_{j, j+1, \ldots j+l}^{(n)}$ are multilinear in the generators~$u_j$ and correspond to the reduced words~(\ref{TL_reduced_word}) in the Temperley-Lieb algebra.

We now proceed with looking for local conserved charges $Q_n$ that commute with the Temperley-Lieb algebraic Floquet evolution operator~$U_F$ in Eq.~(\ref{TL_U_F}). It is convenient to rewrite the integrability condition~$[\,Q_n, U_F\,] = 0$  as
\be \label{U_F_integrability_condition}
	e^{i T_2 H_2} Q_n e^{- i T_2 H_2} = e^{-i T_1 H_1} Q_n e^{i T_1 H_1}\,.
\ee
In order to find the first conserved charge one can simply make the most general ansatz for~$Q_n$, which is homogeneous and consists of terms that are at most bilinear in the generators~$u_j$. Thus, keeping in mind that the terms on even and odd sites play distinct roles [see Eq.~(\ref{U_F_T_matrix})], for the first conserved charge we make the following ansatz:
\be \label{Q1_ansatz_gen}
\begin{aligned}
	Q_1 =& H_1 + a_0 H_2 + \sum_{j=1}^{M-1}\left( b_0 u_{2j} u_{2j+1} + b_1 u_{2j-1} u_{2j}\right)\\
	&+ \sum_{j=1}^{M-1}\left( c_0 u_{2j+1} u_{2j} + c_1 u_{2j} u_{2j-1}\right),
\end{aligned}
\ee
where $H_{1,2}$ are given by Eq.~(\ref{TL_H1_H2}), and $a_0$, $b_{0,1}$, $c_{0,1}$ are yet unknown coefficients to be determined from Eq.~(\ref{U_F_integrability_condition}). Anticipating the result, it is convenient to rewrite the ansatz~(\ref{Q1_ansatz_gen}) as
\be \label{Q1_ansatz}
\begin{aligned}
	Q_1 &= H_1 + a_0 \, H_2  + a_1\, \left[ H_1, H_2 \right] + a_2\, {\cal A}, \\
\end{aligned}
\ee
where from Eqs.~(\ref{TL_rels}) and (\ref{TL_H1_H2}) one has~\cite{Q1_note}
\be \label{TL_H_1_H_2_comm}
	[H_1, H_2] = \sum_{j=1}^{2M - 2}(-1)^{j} \left[ u_j, u_{j+1} \right]
\ee
and  we introduced the operator 
\be \label{calA_operator}
	{\cal A}  = \sum_{j=1}^{2M-2} \left\{ u_j, u_{j+1} \right\},
\ee
with $\{ \cdot, \cdot \}$ being the anticommutator. The coefficients~$a_{1,2}$ in Eq.~(\ref{Q1_ansatz}) are related to those in Eq.~(\ref{Q1_ansatz_gen}) via $b_0 = c_1 = a_2 + a_1$ and  $b_1 = c_0 = a_2 - a_1$.
Even though the number of free parameters in the ansatz~(\ref{Q1_ansatz}) is reduced, detailed analysis shows that Eq.~(\ref{Q1_ansatz_gen}) does not lead to any new solutions to Eq.~(\ref{U_F_integrability_condition}), apart from the one in Eq.~(\ref{Q1_ansatz}).

We then substitute the ansatz~(\ref{Q1_ansatz}) for~$Q_1$ into the integrability condition~(\ref{U_F_integrability_condition}) and check whether it can be satisfied for some values of~$a_k$. Remarkably, one can perform the unitary transformations in Eq.~(\ref{U_F_integrability_condition}) analytically and obtain a closed form expression for~$e^{i T_k H_k} Q_1 e^{- i T_k H_k}$, with $k=1,2$. We discuss the details of this calculation in Appendix~\ref{A:TL_adjoints}.
Using Eqs. (\ref{tildeH_2})\,--\,(\ref{Ad_e_Hm_calA}), we find that condition~(\ref{U_F_integrability_condition}) is satisfied if one has
\be \label{equal_T1_T2}
	T_1 = T_2 = T,
\ee
and the coefficients $a_k$ in Eq.~(\ref{Q1_ansatz}) are given by
\be \label{Q1_ansatz_coeffs}
	a_0 = 1,\;\;  a_1 = \frac{i}{2\beta}\sin \beta T, \;\;  a_2 = -\frac{1}{\beta}\sin^2\frac{\beta T}{2}.
\ee
Thus, the local charge~$Q_1$ in Eq.~(\ref{Q1_ansatz}) with the coefficients~$a_k$ from Eq.~(\ref{Q1_ansatz_coeffs}) provides an {\it exact} conservation law of the Temperley-Lieb algebraic Floquet protocol~ (\ref{TL_U_F}), since it commutes with the one-period evolution operator~$U_F = e^{-i T_2 H_2} e^{-i T_1 H_1}$. We emphasize that since~$Q_1$ exists only when~$T_1 = T_2$, the same condition is required for all higher order conserved charges as well.

Note that we were able to derive~$Q_1$ analytically because the adjoint action~$\exp\{s \, \text{ad}_{H_{1,2}} \}$ on the terms linear and bilinear in the Temperley-Lieb generators~$u_j$ can be calculated in a closed form. Unfortunately, for higher order terms this procedure quickly becomes cumbersome. One can still try to find a few higher order charges with brute force, by simply using an ansatz~(\ref{TL_Qn_ansatz_general}) and requiring that its commutator with the evolution operator in the form~(\ref{U_F_T_matrix}) is zero. Proceeding in this way, setting~$T_1 = T_2 = T$ and using the relations~(\ref{TL_Groebner_def}),~(\ref{TL_Groebner_rels}), for the second conserved charge we obtain
\be \label{Q2}
	Q_2 = Q_2^{(2)}+Q_2^{(3)} + Q_2^{(4)} + Q_2^{(\text{edge})}, 
\ee
where $Q_2^{(n)}$ is a term that contains multilinear products of $n$ generators~$u_j$ and acts in the bulk of the chain, whereas $Q_2^{(\text{edge})}$ is the boundary term, which appears due to the fact that the Temperley-Lieb generators are defined on a chain with open boundary conditions. 
Explicitly, for~$Q_2^{(2)}$ we have
\be
	Q_2^{(2)} = b_{2}  [H_1, H_2] + c_{2} {\cal A},
\ee
where ${\cal A}$ is given by Eq.~(\ref{calA_operator}) and the coefficients are
\be
\begin{aligned}
	b_2 &= \frac{i}{2 \beta }\left(\beta^2 + 2 \cos \beta T\right) \tan\frac{\beta T}{2}, \\
	c_2 &= \frac{1}{2\beta}\left( \beta^2 - 4\sin^2 \frac{\beta T}{2} \right).
\end{aligned}
\ee
The terms $Q_2^{(n)}$, with~$3 \leq n \leq 4$, are more complicated and for the sake of readability we introduce the short-hand notations
\be
	C_j^{-} = [u_j, u_{j+1} ], \qquad C^{+}_j = \{u_j, u_{j+1} \}.
\ee
Then, in terms of the generator we have
\be
\begin{aligned}
	Q_2^{(3)} &= b_3 \sum_{j=1}^{2M-3} (-1)^j \left( [ u_j, C^{+}_{j+1} ] - \{ u_j, C^{-}_{j+1} \} \right) \\
	&\qquad + c_3 \sum_{j=1}^{2M-3} \left[ u_j,  C^{-}_{j+1} \right],
\end{aligned}
\ee
where the coefficients are given by
\be
	b_3 = - \frac{i}{2} \tan\frac{\beta T}{2}, \qquad c_3 = 1.
\ee 
The next term is given by
\be
\begin{aligned}
	Q_2^{(4)} =&  \sum_{j=1}^{2M-4} (-1)^j\Bigl( b_4 [C_j^{-}, C_{j+2}^{-}]  + c_4 [C_j^{+}, C_{j+2}^{+}] \Bigr) \\
	& + d_4 \sum_{j=1}^{2M-4} \left(  [C_j^{-}, C_{j+2}^{+}] + [C_j^{+}, C_{j+2}^{-}]  \right) ,
\end{aligned}
\ee
where the coefficients read as
\be
\begin{aligned}
	&b_4 = \frac{i}{2 \beta} \sin \beta T, \quad d_4 = - \frac{1}{\beta} \sin^2\frac{\beta T}{2}, \\
	&c_4 = - \frac{i}{\beta} \sin^2\frac{\beta T}{2} \tan\frac{\beta T}{2}.
\end{aligned}
\ee
Finally, for the boundary term one has
\be
\begin{aligned}
	Q_2^{(\text{edge})} &= - (u_1 + u_{2M-1}) + b_{\text{edge}} \left( C_1^{+} +C_{2M-2}^{+} \right) \\
	& + c_{\text{edge}} \left( C_1^{-} - C_{2M-2}^{-} \right),
\end{aligned}
\ee
with the constants 
\be
	b_{\text{edge}} = \frac{2}{\beta}\sin^2\frac{\beta T}{2}, \quad c_{\text{edge}} = \frac{i}{\beta} \cos \beta T \tan\frac{\beta T}{2}.
\ee
We have checked that $Q_2$ commutes with both~$Q_1$ in Eq.~(\ref{Q1_ansatz}) and~$U_F$ in Eq.~(\ref{U_F_T_matrix}) for $T_1= T_2 = T$. Note that~$Q_2$ in Eq.~(\ref{Q2}) already includes the terms with up to four generators~$u_j$, even though the previous charge~$Q_1$ in Eq.~(\ref{Q1_ansatz}) contains at most bilinear terms. The reason is that we are dealing with open boundary conditions, and it is well known that in the absence of translational invariance there only exist conserved charges whose maximal support is even. For integrable spin chains with open boundary conditions this is shown in Ref.~\cite{Grabowski1996}.

 We have also obtained the next conserved charge~$Q_3$ and verified that it commutes with~$Q_1$, $Q_2$, and $U_F$ under the same condition. The explicit form of~$Q_3$ is extremely bulky and not illuminating. For this reason, we do not present it here. Expressions for higher order charges~$Q_n$ are even more complicated, in particular because of the boundary terms which proliferate for larger~$n$~\cite{boundaries_note}. Despite the fact that the general form of~$Q_n$ is missing, we strongly believe that it should be possible to obtain it in the closed form. 
We thus conjecture that for~$T_1 = T_2$ the Temperley-Lieb algebraic Floquet protocol in Eq.~(\ref{TL_U_F}) is integrable for arbitrary~$\beta$ and there exists a macroscopically large number of local conserved charges~$Q_n$ that commute with the Floquet evolution operator~$U_F$ and represent the conservation laws of the Floquet~Hamiltonian~$H_F$. 
We leave the proof of our conjecture for future work.
Let us emphasize that integrability of the Temperley-Lieb algebraic Floquet protocol automatically extends to {\it every} representation of the Temperley-Lieb algebra, even to non-Hermitian ones. In particular, the results of this Section cover the protocol considered in Section~\ref{sec:Floquet_integrability} for the $3$-state Potts model, as well as the ones for the TFIM~\cite{Mishra2015} and the Heisenberg model~\cite{Gritsev2008}, since all three models correspond to different representation of the Temperley-Lieb algebra.

\subsection{Three-step stroboscopic Floquet protocol and its integrability} \label{sec:3-step}

One can easily obtain a slightly more general result. Namely, consider the following evolution operator
\be \label{TL_U_F_prime}
	U_F^{\prime}(\lambda) = e^{- i \lambda T H_1} e^{- i T H_2} e^{- i (1-\lambda) T H_1},
\ee
with $\lambda \in {\mathds R}$, which corresponds to a three-step stroboscopic Floquet protocol. On the other hand, $U_F^{\prime}(\lambda)$ in Eq.~(\ref{TL_U_F_prime}) is nothing other than the adjoint action of~$H_1$ on the two-step Floquet evolution operator~$U_F$ from Eq.~(\ref{TL_U_F}) at the integrable point~$T_1 = T_2 = T$, i.e. $U_F^{\prime}(\lambda) = e^{- i \lambda T H_1} U_F e^{i \lambda T H_1}$. Therefore, $U_F^{\prime}$ obviously commutes with an operator 
\be
	Q_n^{\prime}(\lambda) = e^{- i \lambda T H_1} Q_n e^{i \lambda T H_1},
\ee
where~$Q_n$ is the $n$th conserved charge of the two-step protocol~$U_F$ with $T_1 = T_2 = T$.
This means that the three-step Floquet protocol~(\ref{TL_U_F_prime}) is integrable by construction.
Using the results of Appendix~\ref{A:TL_adjoints}, namely Eqs.~(\ref{tildeH_2}), (\ref{Ad_e_Hm_ad_Hm_Hn_1_2}), and (\ref{Ad_e_Hm_calA}), we can immediately find the explicit form of the first non-trivial charge~$Q_1^{\prime}$, which reads 
\begin{widetext}
\be \label{Q1_prime}
\begin{aligned}
	Q_1^{\prime} (\lambda) &= H_1 + H_2 + \frac{i}{\beta}\cos\frac{\beta T}{2} \, \sin\frac{(1-2\lambda)\beta T }{2}\,\left[ H_1, H_2 \right] - \frac{1}{\beta} \left( \sin^2\frac{\lambda \beta T}{2} + \sin^2\frac{(1-\lambda)\beta T}{2} \right) \, {\cal A} \\
	&\frac{4}{\beta^2} \sin\frac{\lambda\beta T}{2}\,\sin\frac{(1-\lambda)\beta T}{2} \left\{  i \sin\frac{(1-2\lambda)\beta T}{2} \, {\cal K}_0 + \cos\frac{(1-2\lambda)\beta T}{2} \, {\cal P}_0 - \cos\frac{\beta T}{2}\, \left( 2H_1 + {\cal R}_0 \right) \right\} \\
	&-\frac{16}{\beta^3} \sin^2\frac{\lambda\beta T}{2}\,\sin^2\frac{(1-\lambda)\beta T}{2} \, {\cal S}_0,
\end{aligned}
\ee
\end{widetext}
where~${\cal A}$ is given by Eq.~(\ref{calA_operator}), and the operators~${\cal K}_0$, ${\cal P}_0$, ${\cal R}_0$, and~${\cal S}_0$ are defined in Eq.~(\ref{adHmHn_ops}).
Taking $\lambda = 0$ in Eq.~(\ref{Q1_prime}) we recover Eq. (\ref{Q1_ansatz}) with the coefficients given by Eq. (\ref{Q1_ansatz_coeffs}), whereas for $\lambda = 1$ we obtain Eq. (\ref{Q1_ansatz}) with $H_1$ and $H_2$ being swapped.  Likewise, one immediately sees that the Floquet protocol 
\be
	U_F^{\prime\prime}(\mu) = e^{- i (1-\mu) T H_2} e^{- i T H_1} e^{ i \mu T H_2},
\ee
with $\mu \in {\mathds R}$ is also integrable by construction, with the charges given by $Q_n^{\prime\prime}(\mu) = e^{i \mu T H_2} Q_n e^{- i \mu T H_2}$. Explicit form of~$Q_1^{\prime\prime}(\mu)$ follows from Eqs.~(\ref{tildeH_2}), (\ref{Ad_e_Hm_ad_Hm_Hn_1_2}), and (\ref{Ad_e_Hm_calA}).


\section{Discussion and conclusions}\label{sec:Conclusion}

Here we summarize the main results of the present work and formulate some open questions for future research.
In the first part of the paper (Section~\ref{sec:Floquet_protocol}) we have proposed a stroboscopic Floquet protocol for generating very simple albeit non-local pairwise entanglement between distant qutrits in the 1D $3$-state Potts model with periodically kicked transverse field. We consider a realistic and experimentally relevant case of a finite chain with~$2N$ sites and open boundary conditions. 
The protocol consists of two main stages. At the first stage we perform a state preparation procedure and transform an initial polarized state of the chain (i.e. every qutrit is in one and the same internal state) into a bipartire state in which the two halves of the chain are completely isolated but each of them is highly entangled. At the second stage of the protocol we iteratively eliminate the entanglement {\it inside} the left and right halves of the chain and at the same time create a simple but non-local pairwise entanglement {\it between} them. At the end of the second stage the system is in product state of maximally entangled non-local Bell-like qutrit pairs. The protocol is illustrated in Fig.~\ref{F:Potts_entanglement_gen}. Note that the protocol requires tuning the transverse field switching frequency to a specific value. We argue that the reason for this condition is deeply rooted into the Floquet integrability of the protocol.

The second part of the paper (Section~\ref{sec:Floquet_integrability}) is dedicated to the idea of Floquet integrability, which is understood as the presence of local conserved charges that commute with the Floquet evolution operator~$U_F$ and, consequently, with an effective Floquet Hamiltonian~$H_F$.
Motivated by the fact that the $3$-state Potts model can be thought of as a representation of the Temperley-Lieb algebra, which has remarkably many different representations corresponding to other paramount physical models, we rewrite the stroboscopic two step Floquet protocol in terms of the Temperley-Lieb algebra generators. We then find the first two non-trivial conservation laws of the Temperley-Lieb algebraic Floquet protocol, and explicitly construct the corresponding conserved charges. We then conjecture that the general closed form expression for the conserved charges can be found, although the proof of our conjecture is beyond the scope of this work.

\section*{Acknowledgments} 
We thank D.~Abanin, E.~Demler, O.~Gamayun, M.~Lukin, Y.~Miao, and A.~Polkovnikov for useful discussions. The results of D.V.K. on the development of the protocol for generating non-local entangled qutrit pairs and the Floquet integrability of the protocol were supported by the Russian Science Foundation Grant No. 19-71-10091 
(parts of Sec.~\ref{sec:Floquet_protocol} and Sec.~\ref{sec:Floquet_integrability}). 
Analysis of the three-step Floquet protocol (Sec. IVc) has been completed with the support of the Leading Research Center on Quantum Computing Program (Agreement No. 014/20). The work of A.K.F. on the study of the three-state Potts model (in particular, Sec. II) has been supported by the Russian Science Foundation Grant No. 19-71-10092.

\appendix
\begin{widetext}
\section{Derivation of $V_0$ in Eq.~(\ref{V_0_exp_series})} \label{A:V0_exp_series}

The easiest way to derive Eq.~(\ref{V_0_exp_series}) is by using the representation~(\ref{TL_Potts_rep}) of the Temperley-Lieb algebra:
\be \label{V_0_Potts_TL_rep}
	u_{2N} = \frac{1}{\sqrt{3}} \left( {\mathds 1} +  X_N^{\dag} X_{N+1} + X_N X_{N+1}^{\dag} \right).
\ee
 We then write~$V_0$ from Eq.~(\ref{V_0_def}) as
\be
	V_0  = \exp \left\{ i J T \, \left( X_N^{\dag} X_{N+1} + X_N X_{N+1}^{\dag} \right) \right\} = e^{- i J T } e^{\sqrt{3} i J T u_{2N}}.
\ee
Taking into account that~$u_{2N}^2 = \sqrt{3} u_{2N}$ according to Eq.~(\ref{TL_rels}), we immediately obtain
\be
	V_0  = e^{-  i TJ } \left( {\mathds 1} + \sum_{k = 1}^{+\infty} \frac{(\sqrt{3} i J T)^k}{k!} 3^{(k-1)/2} u_{2N} \right) = e^{-i TJ } \left[ {\mathds 1} + \frac{1}{\sqrt{3}}  \left(e^{3 i J T} - 1 \right) u_{2N} \right].
\ee
Using~Eq.~(\ref{V_0_Potts_TL_rep}) for~$u_{2N}$ we arrive at
\be
	V_0 = \mu {\mathds 1} + \nu \left( X_N^{\dag} X_{N+1} + X_N X_{N+1}^{\dag} \right), \quad \mu = \nu + e^{-i J T}, \quad \nu = \frac{1}{3} e^{-i J T} \left( e^{3 i J T} - 1 \right).
\ee
We then write~$X_j = {\cal J}_j^{(1,0)}$ and $X_j^{\dag} = X_j^2 = {\cal J}_j^{(2,0)}$, as follows from Eq.~(\ref{Z_X_to_J}), and obtain Eq.~(\ref{V_0_exp_series}) of the main text.

\section{Generalized hyperbolic functions}
In this Appendix we review basic properties of the generalized hyperbolic functions, and present some relations that are useful for our purposes. We closely follow the discussion in Ref.~\cite{Ungar1982}. Generalized hyperbolic functions~$H_{n,k}(x)$ of order~$n$ and $k$-th kind are solutions to an ordinary differential equation 
\be
	\frac{d^n}{d z^n}H_{n,k}(z) = H_{n,k}(z), \qquad 0 \leq k \leq n-1.
\ee
They have the following series representation:
\be \label{gen_hyperbolic_f_series}
	H_{n,k} (z) = \sum_{r=0}^{+\infty} \frac{z^{n r + k} }{(n r + k)!}, \qquad z \in {\mathbb C},
\ee
from which one immediately obtains
\be
	\frac{d}{d z} H_{n,k}(z) = H_{n,k-1}(z), \qquad H_{n,-1} = H_{n,n-1}(z).
\ee
Clearly, Eq.~(\ref{gen_hyperbolic_f_series}) for $n=2$ reduces to the series for usual hyperbolic functions, i.e. $H_{2,0} = \cosh z$ and $H_{2,1} = \sinh z$. We are interested in the case~$n=3$, for which Eq.~(\ref{gen_hyperbolic_f_series}) gives
\be \label{gen_hyperbolic_f_explicit}
\begin{aligned}
	H_{3,0}(z) & \equiv h_0(z) =  \frac{1}{3}\left[ e^z + 2 e^{-z/2} \cos\frac{\sqrt{3}z}{2} \right] 
		= \frac{1}{3} \left( e^z + e^{\omega z} + e^{\omega^2 z} \right),\\
	H_{3,1}(z) & \equiv h_1(z) =  \frac{1}{3}\left[ e^z - 2 e^{-z/2} \cos\left(\frac{\sqrt{3}z}{2} + \frac{\pi}{3} \right) \right] 
		= \frac{1}{3} \left( e^z + \omega e^{\omega z} + \omega^2 e^{\omega^2 z} \right),\\
	H_{3,2}(z) & \equiv h_2(z) =  \frac{1}{3}\left[ e^x - 2 e^{-x/2} \cos\left(\frac{\sqrt{3}x}{2} - \frac{\pi}{3} \right) \right]
		= \frac{1}{3} \left( e^z + \omega^2 e^{\omega z} + \omega e^{\omega^2 z} \right),\\
\end{aligned}
\ee
where we introduced the functions~$h_k(z)$ for brevity and $\omega = e^{2\pi i /3}$, as in the rest of the paper. 
One immediately sees that~$h_k(z)$ satisfy
\be \label{gen_hyperbolic_sym}
\begin{aligned}
	&h_0(\omega z) = h_0(z),  		&& h_0(\omega^2 z) = h_0(z), \\
	&h_1(\omega z) =  \omega ^2 h_1(z),  && h_1(\omega^2 z) = \omega h_1 (z) \\
	&h_2(\omega z) = \omega h_2 (z) && h_2(\omega^2 z) = \omega^2 h_2(z),\\
\end{aligned}
\ee
 and one has~$e^z = \sum_{l=0}^2 h_l(z)$. 
Let us combine~$h_k(z)$ into a circulant matrix 
\be 
	{\mathds H}(z)= \begin{pmatrix}
					h_0(z) & h_2(z)& h_1(z)\\
					h_1(z) & h_0(z)& h_2(z)\\
					h_2(z) & h_1(z)& h_0(z)\\
				\end{pmatrix}.
\ee
One can show that the matrix~${\mathds H}(z)$ satisfies~$\det {\mathds H}(z) = 1$ and the following group property:
\be \label{H_group_prop}
	{\mathds H}(z_1) {\mathds H}(z_2) = {\mathds H}(z_2) {\mathds H}(z_1) = {\mathds H}(z_1 + z_2).
\ee
Thus, for the generalized hyperbolic functions of order~$3$ we have
\be
\begin{aligned}
	&h_0(z_1 + z_2) = h_0(z_1) h_0(z_2) + h_1(z_1) h_2(z_2) + h_2(z_1) h_1(z_2),\\
	&h_1(z_1 + z_2) = h_0(z_1) h_1(z_2) + h_1(z_1) h_0(z_2) + h_2(z_1) h_2(z_2),\\
	&h_2(z_1 + z_2) = h_0(z_1) h_2(z_2) + h_1(z_1) h_1(z_2) + h_2(z_1) h_0(z_2).
\end{aligned}
\ee
The symmetry relations~(\ref{gen_hyperbolic_sym}) can be written in a compact form
\be \label{gen_hyperbolic_sym_matrix}
	{\mathds H}(\omega^m z) = Z^{-m} {\mathds H}(z) Z^m,
\ee
where $0 \leq m \leq 2$ and the matrix $Z  = \text{diag}\{ 1, \omega, \omega^2 \}$ coincides with the one given in Eq.~(\ref{Z_X_matrix_rep}).
One can also easily check the following interesting relations:
\be \label{h_summation}
\begin{aligned}
	&\sum_{l = 0}^2 h_l(x) h_l(y) = \frac{1}{3}\left[ e^{x+y} + 2 e^{-(x+y)/2} \cos\frac{\sqrt{3}(x-y)}{2} \right],\\
	&\sum_{l = 0}^2 h_{l+k \,\text{mod}\,3}(x) h_l(y) = \frac{1}{3}\left[ e^{x+y} - 2 e^{-(x+y)/2} \cos\left( \frac{\sqrt{3}(x-y)}{2} +(-1)^{k-1} \frac{\pi}{3}\right)\right], \qquad k=1,2.
\end{aligned}
\ee
Finally, we note that~Eqs.~(\ref{H_group_prop}) [with $n \times n$ circulant matrix] and~(\ref{gen_hyperbolic_sym_matrix}) [with $0\leq m \leq n-1$] also hold for the generalized hyperbolic functions of order $n$~\cite{Ungar1982, Muldoon1996, Muldoon2005}, and all other relations can be easily extended to the case of arbitrary~$n$.

\section{Adjoint actions} \label{A:adjoints}

In this Appendix we present a detailed derivation of the adjoint actions 
$
e^{i \alpha  \, \text{ad}_{ {\cal J}_j^{ \boldsymbol{m} } }}  {\cal J}_j^{\boldsymbol{p}}, 
$ and 
$
e^{i \alpha  \, 
	\text{ad}_{ 
	 {\cal J}_{j + \ell}^{{\boldsymbol m}} {\cal J}_{j+ \ell + 1}^{{\boldsymbol n}} 
	 }
	}  {\cal J}_{j}^{{\boldsymbol p}} {\cal J}_{j + 1}^{{\boldsymbol q}},
$  
which we then use in Appendix~\ref{A:V1_V2_Vk_derivation} to obtain the explicit forms for~$V_1$, $V_2$, and~$V_k$.

\subsection{Adjoint action of ${\cal J}_j^{(m_1, m_2)}$ on ${\cal J}_j^{(p_1, p_2)}$ }

Using Eq.~\eqref{J_comm_rel} for the commutator $[ {\cal J}_j^{{\boldsymbol m}}, {\cal J}_j^{{\boldsymbol p}} ] \equiv \text{ad}_{ {\cal J}_j^{\boldsymbol{m} } } {\cal J}_j^{\boldsymbol{p}} $ multiple times, we obtain:
\be
	\text{ad}^k_{ {\cal J}_j^{\boldsymbol{m} } } {\cal J}_j^{\boldsymbol{p}}  = (-2i)^k \sin^k \Bigl( \frac{2\pi}{3} \, \boldsymbol{m} \times \boldsymbol{p} \Bigr) {\cal J}_j^{ \boldsymbol{p} + k\, \boldsymbol{m} },
\ee
where the components of the two-dimensional vector $\boldsymbol{p} + k\, \boldsymbol{m}$ are $\text{mod }3$. Therefore, we have the following adjoint action:
\be \label{magnetic_term_conj}
\begin{aligned}
	e^{i \alpha  \, \text{ad}_{ {\cal J}_j^{ \boldsymbol{m} } }} & {\cal J}_j^{\boldsymbol{p}}   
	= \sum_{n = 0}^{\infty} \frac{(i \alpha)^n}{n!} \text{ad}_{ {\cal J}_j^{ \boldsymbol{m} } }^n {\cal J}_j^{\boldsymbol{p}} 
	 = \sum_{k=0}^2 \sum_{n = 0}^{\infty} \frac{(i \alpha)^{3n+k}}{(3n + k)!} \text{ad}_{ {\cal J}_j^{ \boldsymbol{m} } }^{3n+k} {\cal J}_j^{\boldsymbol{p}} 
	 = \sum_{k=0}^2 h_k \bigl( \alpha \, \xi_{\boldsymbol{m}, \boldsymbol{p}} \bigr) {\cal J}_j^{\boldsymbol{p} + k \boldsymbol{m} },
\end{aligned}
\ee
where we denoted $\xi_{\boldsymbol{m}, \boldsymbol{p}} = 2  \sin \left[ 2\pi \, (\boldsymbol{m} \times \boldsymbol{p}) / 3 \right]$, 
and $h_k(z)$ are the generalized hyperbolic functions of order~$3$ and $k$-th kind, given by Eq.~(\ref{gen_hyperbolic_f_explicit}).

\subsection{Adjoint action of ${\cal J}_{j+\ell}^{(m_1, m_2)} {\cal J}_{j+\ell+1}^{(n_1, n_2)}$ on ${\cal J}_{j}^{(p_1, p_2)} {\cal J}_{j+ 1}^{(q_1, q_2)}$}

We proceed with calculating the adjoint actions containing more than two operators, i.e. the commutators of the form
\be \label{ad_JJ_JJ_gen}
	\text{ad}_{ 
	 {\cal J}_{j+\ell}^{{\boldsymbol m}} {\cal J}_{j+\ell+1}^{{\boldsymbol n}} 
	 } {\cal J}_{j}^{{\boldsymbol p}} {\cal J}_{j + 1}^{{\boldsymbol q}} = \left[ {\cal J}_{j+\ell}^{{\boldsymbol m}} {\cal J}_{j+\ell+1}^{{\boldsymbol n}},  {\cal J}_{j}^{{\boldsymbol p}} {\cal J}_{j + 1}^{{\boldsymbol q}} \right].
\ee
Clearly, the resulting expression differs from zero only if $\ell =0, \pm 1$. Taking $\ell = 0$, Eq.~(\ref{ad_JJ_JJ_gen}) yields
\be \label{ad_JJ_JJ_same_indices}
	\text{ad}_{ 
	 {\cal J}_{j}^{{\boldsymbol m}} {\cal J}_{j+1}^{{\boldsymbol n}} 
	 } {\cal J}_{j}^{{\boldsymbol p}} {\cal J}_{j + 1}^{{\boldsymbol q}} = -2 i \sin \left[ \frac{2\pi}{3} \left( {\boldsymbol m} \times {\boldsymbol p} + {\boldsymbol n} \times {\boldsymbol q} \right) \right] {\cal J}_{j}^{{\boldsymbol p} + {\boldsymbol m}} {\cal J}_{j + 1}^{{\boldsymbol q} + {\boldsymbol n}}, 
\ee
where we took into account that the operators~${\cal J}_j^{{\boldsymbol m}}$ commute on different sites and used Eq.~(\ref{J_J_product}). Thus, for the adjoint action we obtain
\be \label{exp_ad_JJ_same_indices}
	e^{i \alpha  \, 
	\text{ad}_{ 
	 {\cal J}_{j}^{{\boldsymbol m}} {\cal J}_{j+1}^{{\boldsymbol n}} 
	 }
	}  {\cal J}_{j}^{{\boldsymbol p}} {\cal J}_{j + 1}^{{\boldsymbol q}}
	= \sum_{k=0}^2 h_k( \alpha \zeta_{ \boldsymbol{m}, \boldsymbol{p} }^{ { \boldsymbol{n}, \boldsymbol{q} } }) {\cal J}_{j}^{{\boldsymbol p} + k {\boldsymbol m}} {\cal J}_{j + 1}^{{\boldsymbol q} + k {\boldsymbol n}},
\ee
where $\zeta_{ \boldsymbol{m}, \boldsymbol{p} }^{ { \boldsymbol{n}, \boldsymbol{q} } } = 2 \sin[ 2\pi ( {\boldsymbol m} \times {\boldsymbol p} + {\boldsymbol n} \times {\boldsymbol q} ) /3]$. Note that by taking~${\boldsymbol n} = {\boldsymbol q} = (0,0)$ or ${\boldsymbol m} = {\boldsymbol p} = (0,0)$, we simply reduce Eq.~(\ref{exp_ad_JJ_same_indices}) to~Eq.~(\ref{magnetic_term_conj}).

The result for $\ell = \pm1$ can be easily found in a similar way. In this case one has
\be \label{ad_JJ_pm1}
\begin{aligned}
		& e^{i \alpha  \, 
	\text{ad}_{ 
	 {\cal J}_{j + 1}^{{\boldsymbol m}} {\cal J}_{j+2}^{{\boldsymbol n}} 
	 }
	}  {\cal J}_{j}^{{\boldsymbol p}} {\cal J}_{j + 1}^{{\boldsymbol q}}
	&&=&& \sum_{k=0}^2 h_k( \alpha \xi_{ \boldsymbol{m}, \boldsymbol{q} }) {\cal J}_{j}^{{\boldsymbol p}} {\cal J}_{j + 1}^{{\boldsymbol q} + k {\boldsymbol m}} {\cal J}_{j + 2}^{ k {\boldsymbol n}}, \\
	&e^{i \alpha  \, 
	\text{ad}_{ 
	 {\cal J}_{j - 1}^{{\boldsymbol m}} {\cal J}_{j}^{{\boldsymbol n}} 
	 }
	}  {\cal J}_{j}^{{\boldsymbol p}} {\cal J}_{j + 1}^{{\boldsymbol q}}
	&&=&& \sum_{k=0}^2 h_k( \alpha \xi_{ \boldsymbol{n}, \boldsymbol{p} } ) {\cal J}_{j-1}^{k {\boldsymbol m}} {\cal J}_{j}^{{\boldsymbol p} + k {\boldsymbol n}} {\cal J}_{j + 1}^{ {\boldsymbol q}},
\end{aligned}
\ee
with~$\xi_{{\boldsymbol r}, {\boldsymbol s}}$ given after~Eq.~(\ref{magnetic_term_conj}).

\section{Derivation of~$V_1$ in Eq.~(\ref{V_1_res_J_m}),  $V_2$ in Eq. (\ref{V_2_res_J}), and $V_k$ in Eq. (\ref{V_k_res_J})}
\label{A:V1_V2_Vk_derivation}

In this Appendix we present a detailed derivations of the operators~$V_1$, $V_2$, and~$V_k$. The most important relations obtained this Appendix are summarized in~\ref{A:V1_V2_Vk_derivation_Summary}.

\subsection{Adjoint action of $H_2$ on ${\cal J}_j^{(p_1, p_2)}$ and expression for~$V_1$ in Eq.~(\ref{V_1_res_J_m})}

In order to explicitly calculate~$V_1$ in Eq.~(\ref{V_1}),
let us consider the adjoint action of the transverse field~$H_2$, given by~Eq.~(\ref{Potts_H1_H2_J}), on ${\cal J}_j^{{\boldsymbol p}}$. Using Eq.~(\ref{magnetic_term_conj}), we obtain
\be \label{ad_H_2_J}
	e^{i \alpha \, \text{ad}_{H_2} } {\cal J}_j^{{\boldsymbol p}} = \prod_{n=1}^2 e^{i \alpha  \, \text{ad}_{ {\cal J}_j^{ (0,n) } }} {\cal J}_j^{{\boldsymbol p}} 
	= \prod_{n=1}^{2} \sum_{k=0}^{\infty} \frac{(i f T)^k}{k!} \text{ad}_{ {\cal J}_j^{(0,n)} }^k {\cal J}_j^{{\boldsymbol p}} = \sum_{k,q=0}^2 h_k \bigl( \alpha \, \xi_{(0,1), \boldsymbol{p}} \bigr) h_q \bigl( \alpha \, \xi_{(0,2), \boldsymbol{p}} \bigr) {\cal J}_j^{ (p_1, p_2 + k +2 q) },
\ee
where $\xi_{(0,n), \boldsymbol{p}} = -2\sin(2 n p_1 \pi/3)$ and $1 \leq j \leq 2N$. Keeping in mind that~$p_2 + k + 2q$ should be taken $\text{mod }3$, we can rewrite Eq.~(\ref{ad_H_2_J}) as 
\be \label{ad_H_2_J_G}
	\prod_{n=1}^2 e^{i \alpha  \, \text{ad}_{ {\cal J}_j^{ (0,n) } }} {\cal J}_j^{{\boldsymbol p}} = \sum_{k=0}^{2} G_k^{(1,2)}(- \alpha, p_1) {\cal J}_j^{ (p_1, p_2+k) }.
\ee
In deriving Eq.~(\ref{ad_H_2_J_G}) we used the explicit expressions $\xi_{(0,m), {\boldsymbol p}} = - 2 \sin [2 m p_1 \pi /3 ]$ and introduced the function
\be \label{G_function_def}
	G_k^{(m, n)}(\alpha, p ) \equiv \sum_{l=0}^{2}h_{l+k \, \text{mod}\, 3} \bigl( 2\alpha \sin[ 2 m p \pi /3 ] \bigr) \, h_l \bigl( 2\alpha \sin[ 2 n p \pi /3 ] \bigr).
\ee
Taking into account that~$p \in \{ 0,1,2\}$ and using the relations~(\ref{h_summation}), we obtain
\be \label{G_function_res}
\begin{aligned}
	G_k^{(1,2)}(\alpha, p) =& \left\{ \delta_{p,0} + \frac{1}{3} \left( \delta_{p,1} + \delta_{p,2} \right) \left( 1 + 2 \cos 3\alpha \right) \right\} \delta_{k,0} \\
	&+ \frac{1}{3} \left\{ \delta_{p,1} \left[ 1 + 2 \cos \left(3\alpha - \frac{2\pi}{3}\right) \right] + \delta_{p,2} \left[ 1 + 2 \cos \left(3\alpha + \frac{2\pi}{3}\right) \right] \right\} \delta_{k,1} \\
	&+ \frac{1}{3} \left\{ \delta_{p,1} \left[ 1 + 2 \cos \left(3\alpha + \frac{2\pi}{3}\right) \right] + \delta_{p,2} \left[ 1 + 2 \cos \left(3\alpha - \frac{2\pi}{3}\right) \right] \right\} \delta_{k,2}\\
	=&\delta_{p,0} \delta_{k,0} + \frac{1}{3}\left( 1-\delta_{p,0}\right) \left[ 1 + 2 \cos \left(3\alpha - \frac{2 p k \pi}{3}\right)\right]
\end{aligned}
\ee	
Then, we further simplify the adjoint action in Eq.~(\ref{ad_H_2_J_G}) and it becomes
\be
	\prod_{n=1}^2 e^{i \alpha  \, \text{ad}_{ {\cal J}_j^{ (0,n) } }} {\cal J}_j^{(p_1, p_2)} = \delta_{p_1,0} {\cal J}_j^{(p_1, p_2)} + \frac{1}{3}\left( 1-\delta_{p_1,0}\right) \sum_{k=0}^{2} \left[ 1 + 2 \cos \left(3\alpha + \frac{2 p_1 k \pi}{3}\right)\right] {\cal J}_j^{ (p_1, p_2+k) }.
\ee
Changing the summation index to $p_1 q$ and taking into account that for~$p_1 \in \{1,2\}$ one has $p_1^2 \,\text{mod}\, 3 = 1$, we write
\be \label{ad_H_2_res_gen}
	e^{i \alpha \, \text{ad}_{H_2} } {\cal J}_j^{(p_1, p_2)}  = \delta_{p_1,0} {\cal J}_j^{(0, p_2)} + \frac{1}{3}\left( 1-\delta_{p_1,0}\right) \sum_{k=0}^{2} \left[ 1 + 2 \cos \left(3\alpha + \frac{2 k \pi}{3}\right)\right] {\cal J}_j^{ (p_1, p_2 + k p_1) }.
\ee
It is now clear that by choosing
\be \label{alpha_simplification}
	\alpha = \alpha_m = \frac{2\pi}{9} (3 l - m), \quad\text{with} \quad l \in {\mathbb Z}, \; m \in \{ 0,1,2\},
\ee
the expression~(\ref{ad_H_2_res_gen}) simplifies, since in the sum over~$k$ only the term with $k = m$ differs from zero. We then have
\be \label{ad_H_2_J_final_res}
	e^{i \alpha_m \, \text{ad}_{H_2} } {\cal J}_j^{(p_1, p_2)} = \delta_{p_1,0} {\cal J}_j^{(0, p_2)} + \left( 1-\delta_{p_1,0}\right) {\cal J}_j^{(p_1, p_2 + m p_1)}, \qquad 1 \leq j \leq 2N.
\ee 
Thus, using Eq.~(\ref{V_0_def}) for~$V_0$ and the definition of $V_1$ in Eq.~(\ref{V_1}), for $f T = \alpha_m$ we obtain
\be \label{V_1_FINAL_RES}
	V_1 = e^{i \alpha_m \, \text{ad}_{H_2}  } V_0 = \mu {\mathds 1} + \nu \left( {\cal J}_N^{(1,m)} {\cal J}_{N+1}^{(2,2m)} +  {\cal J}_N^{(2, 2m)} {\cal J}_{N+1}^{(1, m)}\right),
\ee 
which is the expression for $V_1$ in Eq.~(\ref{V_1_res_J_m}). Clearly, $V_1$ produces a maximally entangled two-qutrit state on the sites~$N$ and $N+1$.

\subsection{Adjoint action of $\tilde H_1$ on ${\cal J}_{j}^{(p_1, p_2)}$ with $j =  1, N, N+1, 2N$, and expression for~$V_2$ in Eq.~(\ref{V_2_res_J})}

Let us now proceed with showing that the explicit form of~$V_2$ is given by Eq.~(\ref{V_2_res_J}). To do so, we consider the adjoint action
\be \label{ad_tilde_H_1_J_center}
\begin{aligned}
	& e^{ i \alpha \, \text{ad}_{\tilde H_1}  } {\cal J}_{N}^{{\boldsymbol p}}  &&=&& 
	\exp\left\{ i \alpha \, \text{ad}_{ {\cal J}_{N-1}^{(2,0)} {\cal J}_{N}^{(1,0)} } \right\} 
	\exp\left\{ i \alpha \, \text{ad}_{ {\cal J}_{N-1}^{(1,0)} {\cal J}_{N}^{(2,0)} } \right\}
	 {\cal J}_{N}^{{\boldsymbol p}}, \\
	& e^{ i \alpha \, \text{ad}_{\tilde H_1}  } {\cal J}_{N+1}^{{\boldsymbol p}}  &&=&&
	 \exp\left\{ i \alpha \, \text{ad}_{ {\cal J}_{N+1}^{(2,0)} {\cal J}_{N+2}^{(1,0)} } \right\} 
	 \exp\left\{ i \alpha \, \text{ad}_{ {\cal J}_{N+1}^{(1,0)} {\cal J}_{N+2}^{(2,0)} } \right\}
	 {\cal J}_{N+1}^{{\boldsymbol p}},
\end{aligned}
\ee
where we took into account that~$\tilde H_1$, given by Eqs.~(\ref{Potts_H1_H2_J}) and (\ref{tilde_H1}), does not contain the terms that act on the central link between the sites~$N$ and $N+1$.  Using Eq.~(\ref{ad_JJ_pm1}) with ${\boldsymbol q} = (0,0)$ and~Eq.~(\ref{exp_ad_JJ_same_indices}), from the first line of Eq.~(\ref{ad_tilde_H_1_J_center}) we have
\be \label{ad_tilde_H_1_J_center_G}
\begin{aligned}
	& e^{ i \alpha \, \text{ad}_{ {\cal J}_{N-1}^{(2,0)}{\cal J}_{N}^{(1,0)} } } 
	e^{ i \alpha \, \text{ad}_{ {\cal J}_{N-1}^{(1,0)}{\cal J}_{N}^{(2,0)} } } {\cal J}_N^{{\boldsymbol p}} = 
	e^{ i \alpha \, \text{ad}_{ {\cal J}_{N-1}^{(2,0)}{\cal J}_{N}^{(1,0)} } } \sum_{k=0}^2 h_k\left( \alpha \xi_{(2,0), {\boldsymbol p} } \right) {\cal J}_{N-1}^{ (k,0) } {\cal J}_N^{ (p_1 + 2 k, p_2) } \\
	&  = \sum_{k,q=0}^2 h_k\left( \alpha \xi_{(2,0), {\boldsymbol p} } \right) h_q\left( \alpha \zeta_{(2,0), (k,0) }^{(1,0), (p_1 + 2k, p_2)} \right) {\cal J}_{N-1}^{ (k + 2 q,0) } {\cal J}_N^{ (p_1 + 2 k + q, p_2) }  
	  = \sum_{k=0}^2 G_k^{(1, 2)}(\alpha, p_2) {\cal J}_{N-1}^{(2 k, 0)} {\cal J}_N^{(p_1 + k, p_2)},
\end{aligned}
\ee
where we used the fact that $k + 2 q$  and $p_1 + 2 k + q$ are $\text{mod}\, 3$, along with the expressions $\xi_{(2,0), {\boldsymbol p} } = 2 \sin( 4 p_2 \pi /3)$ and $\zeta_{(2,0), (k,0) }^{(1,0), (p_1 + 2k, p_2)} = 2 \sin[ 2 p_2 \pi /3 ]$, and took into account Eq.~(\ref{G_function_def}) for the function~$G_k^{(m,n)}(\alpha, p)$. Then, using~Eq.~(\ref{G_function_res}) and repeating the steps leading to Eq.~(\ref{ad_H_2_res_gen}), we immediately obtain
\be \label{ad_tilde_H_1_J_central_res_gen_1} 
	e^{ i \alpha \, \text{ad}_{\tilde H_1}  } {\cal J}_{N}^{{\boldsymbol p}}  = \delta_{p_2,0} {\cal J}_N^{(p_1, 0)} + \frac{1}{3}\left( 1-\delta_{p_2,0}\right) \sum_{k=0}^{2} \left[ 1 + 2 \cos \left(3\alpha - \frac{2 k \pi}{3}\right)\right] \, {\cal J}_{N-1}^{(2 k p_2, 0 )} \, {\cal J}_N^{ (p_1 + k p_2, p_2) }.
\ee
It is easy to see that one similarly has
\be \label{ad_tilde_H_1_J_central_res_gen_2} 
	e^{ i \alpha \, \text{ad}_{\tilde H_1}  } {\cal J}_{N+1}^{{\boldsymbol p}} = \delta_{p_2,0} {\cal J}_{N+1}^{(p_1, 0)} + \frac{1}{3}\left( 1-\delta_{p_2,0}\right) \sum_{k=0}^{2} \left[ 1 + 2 \cos \left(3\alpha - \frac{2 k \pi}{3}\right)\right]\, {\cal J}_{N+1}^{ (p_1 + k p_2, p_2) } \, {\cal J}_{N+2}^{(2 k p_2, 0 )}.
\ee
Just like with Eq.~(\ref{ad_H_2_res_gen}), we can simplify Eqs.~(\ref{ad_tilde_H_1_J_central_res_gen_1}) and~(\ref{ad_tilde_H_1_J_central_res_gen_2}) by using~$\alpha = \alpha_m$ from Eq.~(\ref{alpha_simplification}). In this case only the term with $k = 2 m$ does not vanish in Eqs. (\ref{ad_tilde_H_1_J_central_res_gen_1}), (\ref{ad_tilde_H_1_J_central_res_gen_1}), and one obtains
\be \label{ad_tilde_H_1_J_central_final_res}
\begin{aligned} 
	& e^{ i \alpha_m \, \text{ad}_{\tilde H_1}  } {\cal J}_{N}^{{\boldsymbol p}}  &&=&& \delta_{p_2,0} {\cal J}_N^{(p_1, 0)} + \left( 1-\delta_{p_2,0}\right) {\cal J}_{N-1}^{(m p_2, 0 )} \, {\cal J}_N^{ (p_1 + 2 m p_2, p_2) }, \\
	& e^{ i \alpha_m \, \text{ad}_{\tilde H_1}  } {\cal J}_{N+1}^{{\boldsymbol p}}  &&=&& \delta_{p_2,0} {\cal J}_N^{(p_1, 0)} + \left( 1-\delta_{p_2,0}\right)  {\cal J}_{N+1}^{ (p_1 + 2 m p_2, p_2) } \, {\cal J}_{N+2}^{(m p_2, 0 )}.
\end{aligned}
\ee

Then, in order to find the explicit form of $V_2$ from Eq.~(\ref{V_2}), we need to calculate
\be
	V_2 = \tilde U V_1 \tilde U^{\dag} = e^{i f T \, \text{ad}_{H_2} } e^{i J T \, \text{ad}_{\tilde H_1} } V_1,
\ee
where~$V_1$ is given by Eq.~(\ref{V_1_FINAL_RES}). 
Taking~$J T = \alpha_n$ and $f T = \alpha_s$, with $\alpha_{m}$ given by Eq.~(\ref{alpha_simplification}), we obtain
\be
\begin{aligned}
	& e^{i \alpha_s \, \text{ad}_{H_2} } e^{i \alpha_n \, \text{ad}_{\tilde H_1} } {\cal J}_N^{(1,m)} {\cal J}_{N+1}^{(2,2m)} = e^{i \alpha_s \, \text{ad}_{H_2} } {\cal J}_{N-1}^{(n m, 0)}  {\cal J}_N^{(1 + 2 n m, m)} {\cal J}_{N+1}^{(2 + 4 n m, 2m)} {\cal J}_{N+2}^{(2 n m, 0)} \\
	& = {\cal J}_{N-1}^{(n m, s n m)}  {\cal J}_N^{(1 + 2 n m, m + s (1 + 2 n m) )} {\cal J}_{N+1}^{(2 + n m, 2m + s (2 + n m) )} {\cal J}_{N+2}^{(2 n m, 2 s n m)}.
\end{aligned}
\ee
The resulting expression for $V_2$ is especially simple for~$m = n = s$, which corresponds to  $f T = J T = \alpha_m = 2 \pi \left( 3 l - m\right) / 9$. In this case $V_2$ is given by
\be \label{V_2_f=J}
	V_2 = \mu {\mathds 1} + \nu \left( {\cal J}_{N-1}^{(1, m)} {\cal J}_{N}^{(0, m)} {\cal J}_{N + 1}^{(0, 2 m)} {\cal J}_{N + 2}^{(2, 2 m)} + \text{H.c.} \right),
\ee
where we took into account that for $m\in \{ 1, 2\}$ one has $m^2 \,\text{mod}\, 3 = 1$ and $m^3 \,\text{mod}\, 3 = m$. Setting~$l=1$ and $m = 2$ we obtain $f T = J T = 2 \pi / 9$ and $V_2$ reduces to~Eq.~(\ref{V_2_res_J}). Using Eq.~(\ref{Z_X_to_J}), one can rewrite~Eq.~(\ref{V_2_f=J}) as
\be
	V_2 = \mu {\mathds 1} + \nu \left( X_{N-1} Z_{N-1}^m Z_{N}^m Z_{N+1}^{2m} X_{N+2}^2 Z_{N+2}^{2m} + \text{H.c.} \right),
\ee
and we see that $V_2$ entangles the spins on sites $N-1$ and $N+2$, while only changing the {\it phase} on the sites $N$ and $N+1$.
On the other hand, for $n = 2 m$ and $s = m$ [which corresponds to $f T = \alpha_m$ and $J T = \alpha_{2m}$, with $\alpha_n$ given in Eq.~(\ref{alpha_simplification})], we have
\be \label{V_2_J_f=2J}
	V_2 = \mu {\mathds 1} + \nu \left( {\cal J}_{N-1}^{(2, 2m)} {\cal J}_{N}^{(2, 0)} {\cal J}_{N + 1}^{(1, 0)} {\cal J}_{N + 2}^{(1, m)} + \text{H.c.} \right) = \mu {\mathds 1} + \nu \left( X^2_{N-1} Z_{N-1}^{2m} X_{N}^2 X_{N+1} X_{N+2} Z_{N+2}^{m} + \text{H.c.} \right).
\ee
The effect of~$V_2$ in this case is more complicated as compared to Eq.~(\ref{V_2_f=J}), since it now shifts the {\it states} on sites $N$ and $N+1$.

For the adjoint actions $e^{ i \alpha \, \text{ad}_{\tilde H_1}  } {\cal J}_{j}^{{\boldsymbol p}}$ with $j =1$ and $2N$, the result follows immediately from Eqs.~(\ref{ad_tilde_H_1_J_central_res_gen_1}),~(\ref{ad_tilde_H_1_J_central_res_gen_2}).  Replacing~$N+1$ with~$1$ in~Eq.~(\ref{ad_tilde_H_1_J_central_res_gen_2}) gives the expression for~$e^{ i \alpha \, \text{ad}_{\tilde H_1}  } {\cal J}_{1}^{{\boldsymbol p}}$, and changing~$N$ to $2N$ in Eq.~(\ref{ad_tilde_H_1_J_central_res_gen_1}) we obtain the result for~$e^{ i \alpha \, \text{ad}_{\tilde H_1}  } {\cal J}_{2N}^{{\boldsymbol p}}$.

\subsection{Adjoint action of $\tilde H_1$ on ${\cal J}_{j}^{(p_1, p_2)}$ with $j\neq 1, N, N+1, 2N$, and expression for~$V_k$ in Eq.~(\ref{V_k_res_J})}

Finally, we derive the explicit expression for~$V_k$ in Eq.~(\ref{V_k_res_J}). Let us first calculate the following adjoint action, with~$j \neq 1, N, N+1, 2N$:
\be \label{ad_tilde_H_1_gen}
	e^{ i \alpha \, \text{ad}_{\tilde H_1}  } {\cal J}_j^{{\boldsymbol p}} = \exp\left\{ i \alpha \, \text{ad}_{{\cal J}_j^{(2,0)}{\cal J}_{j+1}^{(1,0)} +  {\cal J}_j^{(1,0)}{\cal J}_{j+1}^{(2,0)}}  \right\} \exp\left\{ i \alpha \, \text{ad}_{{\cal J}_{j-1}^{(2,0)}{\cal J}_{j}^{(1,0)} +  {\cal J}_{j-1}^{(1,0)}{\cal J}_{j}^{(2,0)}}  \right\} {\cal J}_j^{{\boldsymbol p}},
\ee
where we used Eqs.~(\ref{Potts_H1_H2_J}), (\ref{tilde_H1}) for~$\tilde H_1$ and took into account that the operators~${\cal J}_j^{(k,0)}$ with $k=1,2$ commute. 
For the first adjoint action in Eq.~(\ref{ad_tilde_H_1_gen}) we can simply use Eq.~(\ref{ad_tilde_H_1_J_center_G}) with~$N$ replaced by~$j$, which yields
\be \label{ad_tilde_H_1_part1}
	e^{ i \alpha \, \text{ad}_{ {\cal J}_{j-1}^{(2,0)}{\cal J}_{j}^{(1,0)} } } 
	e^{ i \alpha \, \text{ad}_{ {\cal J}_{j-1}^{(1,0)}{\cal J}_{j}^{(2,0)} } } {\cal J}_j^{{\boldsymbol p}} 
	 = \sum_{k=0}^2 G_k^{(1, 2)}(\alpha, p_2) {\cal J}_{j-1}^{(2 k, 0)} {\cal J}_j^{(p_1 + k, p_2)}.
\ee
Then, using Eqs.~(\ref{exp_ad_JJ_same_indices}) and (\ref{ad_JJ_pm1}), for the remaining adjoint action in Eq.~(\ref{ad_tilde_H_1_gen}) we obtain
\be \label{ad_tilde_H_1_part2}
\begin{aligned}
	&e^{ i \alpha \, \text{ad}_{ {\cal J}_{j}^{(2,0)}{\cal J}_{j+1}^{(1,0)} } } 
	e^{ i \alpha \, \text{ad}_{ {\cal J}_{j}^{(1,0)} {\cal J}_{j+1}^{(2,0)} } } {\cal J}_{j-1}^{ (2k,0) } {\cal J}_j^{ (p_1 + k, p_2) }  \\
	&\quad= e^{ i \alpha \, \text{ad}_{ {\cal J}_{j}^{(2,0)}{\cal J}_{j+1}^{(1,0)} } } \sum_{r=0}^2 h_r \left( \alpha \xi_{(1,0),(p_1 + k, p_2)} \right) {\cal J}_{j-1}^{ (2k,0) } {\cal J}_j^{ (p_1 + k + r, p_2) } {\cal J}_{j+1}^{(2 r,0)} \\
	&\quad = \sum_{r,s=0}^2 h_r \left( \alpha \xi_{(1,0),(p_1 + k, p_2)} \right) h_s \left( \alpha \zeta_{(2,0),(p_1 + k + r, p_2)}^{(1,0), (2r ,0)} \right) {\cal J}_{j-1}^{ (2k,0) } {\cal J}_j^{ (p_1 + k + r + 2 s, p_2) } {\cal J}_{j+1}^{(2 r + s,0)}\\
	& \quad= \sum_{r=0}^2 G_r^{(1,2)}(\alpha, p_2) {\cal J}_{j-1}^{ (2k,0) } {\cal J}_{j}^{(p_1 +k + r, p_2)} {\cal J}_{j+1}^{(2r,0)},
\end{aligned}
\ee
where we took into account that $\xi_{(1,0),(p_1 + k, p_2)} = 2 \sin[ 2 p_2 \pi/3 ]$ and $\zeta_{(2,0),(p_1 + k + r, p_2)}^{(1,0), (2r ,0)} = 2 \sin[ 4 p_2 \pi / 3 ]$. Thus, combining Eqs.~(\ref{ad_tilde_H_1_gen}),~(\ref{ad_tilde_H_1_part1}), and~(\ref{ad_tilde_H_1_part2}), we have
\be \label{ad_tilde_H_1_res_gen_bulk}
\begin{aligned}
	&e^{ i \alpha \, \text{ad}_{\tilde H_1}  } {\cal J}_j^{{\boldsymbol p}} = \sum_{k,r=0}^2 G_k^{(1, 2)}(\alpha, p_2) \, G_r^{(1,2)}(\alpha, p_2) \; {\cal J}_{j-1}^{ (2k,0) } {\cal J}_{j}^{(p_1 +k + r, p_2)} {\cal J}_{j+1}^{(2r,0)} = \delta_{p_2,0}  {\cal J}_{j}^{(p_1, p_2)}\\
	& \quad + \frac{1}{9} \left( 1 - \delta_{p_2,0} \right) \sum_{k,r=0}^2 \left[ 1 + 2 \cos \left(3\alpha - \frac{2 k \pi}{3}\right)\right] \left[ 1 + 2 \cos \left(3\alpha - \frac{2 r \pi}{3}\right)\right]\; {\cal J}_{j-1}^{ (2k p_2,0) } {\cal J}_{j}^{(p_1 + (r +k )p_2, p_2)} {\cal J}_{j+1}^{(2r p_2,0)},
\end{aligned}
\ee
where we used Eq.~(\ref{G_function_res}) and the fact that~$\delta_{p,0} (1- \delta_{p,0}) \equiv 0$. Clearly, Eq.~(\ref{ad_tilde_H_1_res_gen_bulk}) greatly simplifies if one takes~$\alpha_s$ from~Eq.~(\ref{alpha_simplification}).
In this case the sum over $k$ and $r$ in Eq.~(\ref{ad_tilde_H_1_res_gen_bulk}) contains only one non-zero term corresponding to $k = r = 2s$, and we obtain
\be \label{ad_tilde_H_1_gen_final_res} 
	e^{ i \alpha_s \, \text{ad}_{\tilde H_1}  } {\cal J}_j^{(p_1, p_2)} = \delta_{p_2,0}  {\cal J}_{j}^{(p_1, 0)} + \left( 1 - \delta_{p_2,0} \right) {\cal J}_{j-1}^{ (s \,p_2,0) } {\cal J}_{j}^{(p_1 + s \, p_2, p_2)} {\cal J}_{j+1}^{(s \,p_2,0)}, \qquad j\neq1,N, N+1,2N.
\ee 

Let us now calculate~$V_3$ using the relation
\be \label{V_3_gen_app}
	V_3 = e^{i f T \, \text{ad}_{H_2} } e^{i J T \, \text{ad}_{\tilde H_1} } V_2 
	= e^{i f T \, \text{ad}_{H_2} } e^{i J T \, \text{ad}_{\tilde H_1} } \;\; e^{i f T \, \text{ad}_{H_2} } e^{i J T \, \text{ad}_{\tilde H_1} }  \; \; e^{i f T \, \text{ad}_{H_2} } V_0.
\ee
In what follows we consider the cases $f T = J T = \alpha_m$ and $f T =\alpha_m $, $J T = \alpha_{2m}$ separately. 

\subsubsection{$f T = J T = \alpha_m$} 

We first assume that $f T = J T = \alpha_m$, so that~$V_2$ is given by Eq.~(\ref{V_2_f=J}). Then, we have
\be \label{V3_J_factors}
\begin{aligned}
	& e^{i f T \, \text{ad}_{H_2} } e^{i J T \, \text{ad}_{\tilde H_1} } {\cal J}_{N-1}^{(1,m)}&&\stackrel{(\ref{ad_tilde_H_1_gen_final_res})}{=}&& e^{i f T \, \text{ad}_{H_2} } {\cal J}_{N-2}^{(1,0)}{\cal J}_{N-1}^{(2,m)} {\cal J}_{N}^{(1,0)} 
	&&&\stackrel{(\ref{ad_H_2_J_final_res})}{=}&&& {\cal J}_{N-2}^{(1,m)}{\cal J}_{N-1}^{(2, 0)} {\cal J}_{N}^{(1,m)},\\
	& e^{i f T \, \text{ad}_{H_2} } e^{i J T \, \text{ad}_{\tilde H_1} } {\cal J}_{N}^{(0,m)} &&\stackrel{(\ref{ad_tilde_H_1_J_central_final_res})}{=}&& e^{i f T \, \text{ad}_{H_2} } {\cal J}_{N-1}^{(1,0)} {\cal J}_{N}^{(2,m)} &&&\stackrel{(\ref{ad_H_2_J_final_res})}{=}&&& {\cal J}_{N-1}^{(1, m )} {\cal J}_{N}^{(2,0)} ,\\
	& e^{i f T \, \text{ad}_{H_2} } e^{i J T \, \text{ad}_{\tilde H_1} } {\cal J}_{N+1}^{(0, 2m)} &&\stackrel{(\ref{ad_tilde_H_1_J_central_final_res})}{=}&& e^{i f T \, \text{ad}_{H_2} } {\cal J}_{N+1}^{(1,2m)}{\cal J}_{N+2}^{(2,0)} &&&\stackrel{(\ref{ad_H_2_J_final_res})}{=}&&& {\cal J}_{N+1}^{(1, 0)}{\cal J}_{N+2}^{(2, 2m)} ,\\
	& e^{i f T \, \text{ad}_{H_2} } e^{i J T \, \text{ad}_{\tilde H_1} } {\cal J}_{N+2}^{(0, 2m)} &&\stackrel{(\ref{ad_tilde_H_1_gen_final_res})}{=}&& e^{i f T \, \text{ad}_{H_2} } {\cal J}_{N+1}^{(2,0)} {\cal J}_{N+2}^{(1,2m)}{\cal J}_{N+3}^{(2,0)} 
	&&&\stackrel{(\ref{ad_H_2_J_final_res})}{=}&&& {\cal J}_{N+1}^{(2, 2m)} {\cal J}_{N+2}^{(1,0)}{\cal J}_{N+3}^{(2, 2m)}.
\end{aligned}
\ee
Using Eqs.~(\ref{V_2_f=J}) and~(\ref{V_3_gen_app}) for $V_2$ and $V_3$, correspondingly, and multiplying the results in Eq.~(\ref{V3_J_factors}), we obtain
\be \label{V_3_f=J}
	V_3 = \mu {\mathds 1} + \nu \left( {\cal J}_{N-2}^{(1,m)}{\cal J}_{N-1}^{(0, m)} {\cal J}_{N}^{(0,m)} {\cal J}_{N+1}^{(0, 2m)} {\cal J}_{N+2}^{(0, 2m)}{\cal J}_{N+3}^{(2, 2m)} + \text{H.c.} \right),
\ee
where all factors of~$\omega$, which appear [see Eq.~(\ref{J_J_product})] from the products of two operators on the same site, cancel each other. We see that for $f T = J T = \alpha_m$ the action of $V_3$ is similar to that of $V_2$ from Eq.~(\ref{V_2_f=J}). Namely, $V_3$ entangles the sites $N-2$ and $N+3$, whereas on all sites in between it only rotates the phase.

Then, using Eqs.~(\ref{J_mapping_bulk}) and (\ref{J_mapping_central}), one can easily show by induction that for $f T = J T = \alpha_m$ and $1\leq k < N$ the operator~$V_k$  has the following form:
\be \label{V_k_res_app_f=J}
	V_k = \mu {\mathds 1} + \nu \left( {\cal J}_{N-k+1}^{(1,m)} \prod_{j = k-2}^{0} {\cal J}_{N-j}^{(0, m)} \prod_{j = 1}^{k-1} {\cal J}_{N+j}^{(0, 2m)}  \; {\cal J}_{N+k}^{(2, 2m)} + \text{H.c.} \right).
\ee
Quite remarkably, it only entangles the sites~$N - k + 1$ and $N+k$, while on the rest of the sites $ N - k + 1 < j < N+k$ its effect is a simple phase rotation. Due to this fact, the operator $V_1 V_2 \ldots V_k$ in~Eq.~(\ref{psi_k_gen}) produces an entangled state of a very simple product form, as discussed at the end of Section~\ref{sec:Floquet_protocol}.

\subsubsection{$f T =\alpha_m $, $J T = \alpha_{2m}$}

Let us now investigate what happens if one chooses different values of $f T$ and $JT$, and consider $f T =\alpha_m $, $J T = \alpha_{2m}$. In this case $V_2$ is given by Eq.~(\ref{V_2_J_f=2J}), and instead of Eq.~(\ref{V3_J_factors}) we need the relations
\be \label{V3_J_factors_f=2J}
\begin{aligned}
	& e^{i f T \, \text{ad}_{H_2} } e^{i J T \, \text{ad}_{\tilde H_1} } {\cal J}_{N-1}^{(2, 2m)} &&\stackrel{(\ref{ad_tilde_H_1_gen_final_res})}{=}&& e^{i f T \, \text{ad}_{H_2} } {\cal J}_{N-2}^{(1,0)}{\cal J}_{N-1}^{(0, 2m)} {\cal J}_{N}^{(1,0)} &&&\stackrel{(\ref{ad_H_2_J_final_res})}{=}&&& {\cal J}_{N-2}^{(1,m)}{\cal J}_{N-1}^{(0, 2m)} {\cal J}_{N}^{(1,m)},\\
	& e^{i f T \, \text{ad}_{H_2} } e^{i J T \, \text{ad}_{\tilde H_1} } {\cal J}_{N}^{(2,0)} &&\stackrel{(\ref{ad_tilde_H_1_J_central_final_res})}{=}&& e^{i f T \, \text{ad}_{H_2} } {\cal J}_{N}^{(2,0)} &&&\stackrel{(\ref{ad_H_2_J_final_res})}{=}&&& {\cal J}_{N}^{(2, 2m)} ,\\
	& e^{i f T \, \text{ad}_{H_2} } e^{i J T \, \text{ad}_{\tilde H_1} } {\cal J}_{N+1}^{(1, 0)} &&\stackrel{(\ref{ad_tilde_H_1_J_central_final_res})}{=}&& e^{i f T \, \text{ad}_{H_2} } {\cal J}_{N+1}^{(1, 0)} &&&\stackrel{(\ref{ad_H_2_J_final_res})}{=}&&& {\cal J}_{N+1}^{(1, m)},\\
	& e^{i f T \, \text{ad}_{H_2} } e^{i J T \, \text{ad}_{\tilde H_1} } {\cal J}_{N+2}^{(1, m)} &&\stackrel{(\ref{ad_tilde_H_1_gen_final_res})}{=}&& e^{i f T \, \text{ad}_{H_2} } {\cal J}_{N+1}^{(2,0)} {\cal J}_{N+2}^{(0, m)}{\cal J}_{N+3}^{(2,0)} &&&\stackrel{(\ref{ad_H_2_J_final_res})}{=}&&& {\cal J}_{N+1}^{(2, 2m)} {\cal J}_{N+2}^{(0, m)}{\cal J}_{N+3}^{(2, 2m)}.
\end{aligned}
\ee
Using Eq.~(\ref{V_2_J_f=2J}) for $V_2$ and multiplying the terms in Eq.~(\ref{V3_J_factors_f=2J}), from Eq.~(\ref{V_3_gen_app}) for~$V_3$ we obtain
\be \label{V_3_f=2J}
	V_3 = \mu {\mathds 1} + \nu \left( {\cal J}_{N-2}^{(1,m)}{\cal J}_{N-1}^{(0, 2m)}  {\mathds 1}_{N} {\mathds 1}_{N+1} {\cal J}_{N+2}^{(0, m)}{\cal J}_{N+3}^{(2, 2m)} + \text{H.c.} \right).
\ee
At the first sight, the resulting expression for~$V_3$ in Eq.~(\ref{V_3_f=2J}) looks even simpler as compared to Eq.~(\ref{V_3_f=J}). However, the situation turns out to be more complicated, because for $f T =\alpha_m $ and $J T = \alpha_{2m}$, the expression for~$V_k$ strongly depends on~$k$, in contrast to the case of $f T = J T = \alpha_{m}$ [see Eq.~(\ref{V_k_res_app_f=J})]. Indeed, using Eqs.~(\ref{J_mapping_bulk}) and (\ref{J_mapping_central}) it can be easily shown that the next few $V_k$ are given by
\be \label{V_4_5_6_f=2J}
\begin{aligned}
	V_4 &= \mu {\mathds 1} + \nu \left( 
		\mathcal{J}_{N-3}^{(2,2m)} \, \mathcal{J}_{N-2}^{(1,2m)} \, \mathcal{J}_{N-1}^{(0,2m)} \, \mathcal{J}_{N}^{(1,m)} \,\mathcal{J}_{N+1}^{(2,2m)}  \, \mathcal{J}_{N+2}^{(0,m)}  \, \mathcal{J}_{N+3}^{(2,m)}  \, \mathcal{J}_{N+4}^{(1,m)} 
		+ \text{H.c.}
	\right),\\
	V_5 &= \mu {\mathds 1} + \nu \left(
	\mathcal{J}_{N-4}^{(1,m)} \, \mathcal{J}_{N-3}^{(1,0)} \, \mathcal{J}_{N-2}^{(1,0)} \, \mathcal{J}_{N-1}^{(1,0)} \, \mathcal{J}_{N}^{(0,m)} \, \mathcal{J}_{N+1}^{(0,2m)}
    \mathcal{J}_{N+2}^{(2,0)} \, \mathcal{J}_{N+3}^{(2,0)} \, \mathcal{J}_{N+4}^{(2,0)} \, \mathcal{J}_{N+5}^{(2,2m)}
	+ \text{H.c.}
	\right), \\
	V_6 &= \mu {\mathds 1} + \nu \left(
	\mathcal{J}_{N-5}^{(2,2m)} \, \mathcal{J}_{N-4}^{(0,m)}\,  {\mathds 1}_{N-3} \, \mathcal{J}_{N-2}^{(1,m)} \, {\mathds 1}_{N-1} \, \mathcal{J}_{N}^{(1,2m)}
 \,   \mathcal{J}_{N+1}^{(2,m)} \, {\mathds 1}_{N+2} \, \mathcal{J}_{N+3}^{(2,2m)} \,  {\mathds 1}_{N+4} \, \mathcal{J}_{N+5}^{(0,2m)} \,  \mathcal{J}_{N+6}^{(1, m)}
	+ \text{H.c.}
	\right),
\end{aligned}
\ee
where we assumed that $N > 6$ to avoid dealing with the boundaries.
We see that the form of~$V_k$ drastically changes with increasing~$k$. Thus, Eq.~(\ref{V_4_5_6_f=2J}) suggests that in the case $f T =\alpha_m $ and $J T = \alpha_{2m}$ the general expression for~$V_k$ can not be written in a closed form.

Nevertheless, writing~$V_k = \mu {\mathds 1} + \nu ( {\cal V}_k + \text{H.c.} )$ as in Eq.~(\ref{V_k_three_terms}), one can obtain a recursive relation between the indices of ${\cal J}_j^{(p,q)}$ in~${\cal V}_k$ and~${\cal V}_{k+1}$. Indeed, for every~$1 \leq k < N$ we can write~${\cal V}_k$ in the form
\be
	{\cal V}_k = {\cal J}_{N-k+1}^{(p_{N-k+1}, q_{N-k+1})} \ldots {\cal J}_N^{(p_N, q_N)} {\cal J}_{N+1}^{(2 p_{N}, 2 q_{N})}  \ldots {\cal J}_{N+k}^{(2 p_{N-k+1}, 2q_{N-k+1})}.
\ee
Then, for ${\cal V}_{k+1} = \tilde U {\cal V}_k \tilde U =  e^{i f T \, \text{ad}_{H_2} } e^{i J T \, \text{ad}_{\tilde H_1} } {\cal V}_k$ we can write
\be \label{ad_J_string_to_J_string}
\begin{aligned}
	e^{i \alpha_m \, \text{ad}_{H_2} } & e^{i \alpha_{2m} \, \text{ad}_{\tilde H_1} } \;\;  {\cal J}_{N-k+1}^{(p_{N-k+1}, q_{N-k+1})} \ldots {\cal J}_N^{(p_N, q_N)} {\cal J}_{N+1}^{(2 p_{N}, 2 q_{N})}  \ldots {\cal J}_{N+k}^{(2 p_{N-k+1}, 2q_{N-k+1})} \\
	& =  {\cal J}_{N-k}^{(p^{\prime}_{N-k}, q^{\prime}_{N-k})} \ldots {\cal J}_N^{(p^{\prime}_N, q^{\prime}_N)} {\cal J}_{N+1}^{(2 p^{\prime}_{N}, 2 q^{\prime}_{N})}  \ldots {\cal J}_{N+k+1}^{(2 p^{\prime}_{N-k}, 2q^{\prime}_{N-k})}.
\end{aligned}
\ee
The relation~(\ref{ad_J_string_to_J_string}) can be viewed simply as a linear transformation between the indices of~${\cal J}_j^{(p_j, q_j)}$, which can be shown to be 
\begin{equation} \label{J_indices_lin_transform}
\begin{aligned}
	&(p_{N-k}^{\prime}, q_{N-k}^{\prime} ) &&=&& (2m\ q_{N-k+1}, 2\ q_{N-k+1}), \\
	&(q_{N-k+1}^{\prime}, q_{N-k+1}^{\prime} ) &&=&& 
	(2m\ q_{N-k+2} + p_{N-k+1} + 2m\ q_{N-k+1}, 
    2\ q_{N-k+2} +m\ p_{N-k+1} ), \\
	&\qquad\vdots \\
	&(p_{N-j}^{\prime}, q_{N-j}^{\prime} ) &&=&& 
    (2m\ q_{N-j+1} + p_{N-j} + 2m\ q_{N-j} + 2m\ q_{N-j-1},
	 2\ q_{N-j+1} + m\ p_{N-j} + 2\ q_{N-j-1}), \\
	&\qquad\vdots \\
	&(p_{N-1}^{\prime}, q_{N-1}^{\prime} ) &&=&& 
	(2m\ q_{N} + p_{N-1} + 2m\ q_{N-1} + 2m\ q_{N-2}, 
    2\ q_{N} + m\ p_{N-1} + 2\ q_{N-2}), \\ 
	&(p_{N}^{\prime}, q_{N}^{\prime} ) &&=&& 
	(p_{N} + m\ q_{N} + 2m\ q_{N-1}, m\ p_{N} + 2\ q_{N} + 2\ q_{N-1}).
\end{aligned}
\end{equation}
Using Eq.~(\ref{J_indices_lin_transform}) we generate~Fig.~\ref{F:fractal}, which shows that the distributions of~$X_j^p$ and $Z_j^q$ in~${\cal V}_k$ exhibit a fractal structure similar to the Sierpi\'nski carpet.

\subsection{Summary: adjoint action $e^{i \alpha_s \, \text{ad}_{H_2} } e^{i \alpha_m \, \text{ad}_{\tilde H_1} }$ on ${\cal J}_j^{(p_1, p_2)}$} \label{A:V1_V2_Vk_derivation_Summary}
For convenience, here we summarize the most important relations derived in this Appendix. For the values of~$fT= \alpha_s$ and~$JT=\alpha_{m}$, where $\alpha_{r}$ is given in Eq.~(\ref{alpha_simplification}),
the adjoint action~$e^{i \alpha_s \, \text{ad}_{H_2} } e^{i \alpha_m \, \text{ad}_{\tilde H_1} }$ provides the following mapping:
\be \label{J_mapping_bulk}
	e^{i \alpha_m \, \text{ad}_{H_2} } e^{i \alpha_s \, \text{ad}_{\tilde H_1} }: \quad {\cal J}_{j}^{(p_1, p_2)} \to {\cal J}_{j-1}^{(s p_2, m s p_2)}  
		{\cal J}_{j}^{(p_1 + s p_2, (1+ms)p_2 + m p_1 )} 
		{\cal J}_{j+1}^{(s p_2, m s p_2)}, \qquad j \neq 1,N, N+1, 2N,
\ee
and similarly for the boundary and central terms:
\be \label{J_mapping_central}
\begin{aligned}
	& {\cal J}_{j}^{(p_1, p_2)} \to  {\cal J}_{j}^{(p_1 + 2 s p_2, (1 + 2 m s) p_2 + m p_1 )} 
	  {\cal J}_{j+1}^{(s p_2, m s p_2)}, \qquad\qquad j =1, N+1, \\
	& {\cal J}_{j}^{(p_1, p_2)} \to {\cal J}_{j-1}^{(s p_2, m s p_2)} 
	 	{\cal J}_{j}^{(p_1 + 2 s p_2, (1 + 2 m s) p_2 + m p_1 )}, \qquad\qquad j=N,2N. \\ 
\end{aligned}	
\ee 
Eqs.~(\ref{J_mapping_bulk}), (\ref{J_mapping_central}) are easily obtained by combining Eqs.~(\ref{ad_H_2_J_final_res}),~(\ref{ad_tilde_H_1_J_central_final_res}), and~(\ref{ad_tilde_H_1_gen_final_res}).

Choosing $s=m \in \{ 1, 2\}$ and taking into account that $m^2 \,\text{mod}\, 3 = 1$ and $m^3 \,\text{mod}\, 3 = m$, from Eq.~(\ref{J_mapping_bulk}) we obtain
\be \label{J_mapping_bulk_J=f}
	e^{i \alpha_m \, \text{ad}_{H_2} } e^{i \alpha_m \, \text{ad}_{\tilde H_1} }: \quad {\cal J}_{j}^{(p_1, p_2)} \to {\cal J}_{j-1}^{(m p_2, p_2)}  
		{\cal J}_{j}^{(p_1 + m p_2, 2 p_2 + m p_1 )} 
		{\cal J}_{j+1}^{(m p_2, p_2)}, \qquad j \neq 1,N, N+1, 2N,
\ee
whereas Eq.~(\ref{J_mapping_central}) reduces to

\be \label{J_mapping_central_J=f}
\begin{aligned}
\begin{aligned}
	& {\cal J}_{j}^{(p_1, p_2)} \to  {\cal J}_{j}^{(p_1 + 2 m p_2,  m p_1 )} 
	  {\cal J}_{j+1}^{(m p_2, p_2)}, \qquad\qquad j =1, N+1, \\
	&  {\cal J}_{j}^{(p_1, p_2)} \to {\cal J}_{j-1}^{(m p_2, p_2)} 
	 	{\cal J}_{j}^{(p_1 + 2 m p_2, m p_1 )}, \qquad\qquad j=N,2N. \\ 
\end{aligned}	
\end{aligned}	
\ee

\section{Unequal splitting of the chain} \label{A:Floquet_protocol_unequal_split}
In this Appendix we discuss a more general entanglement generating Floquet protocol based on switching off links other than the central one.
Choosing the central link in Eq.~(\ref{tilde_H1}) is not crucial and one can easily switch off any other $M$th link [between sites $M$ and $M+1$], with $1 \leq M \leq 2N-1$. Then, assuming $M \geq 2$ (we are not interesting the case of $M=1$ since the protocol consists of a single period) and taking {\it the same} value for $f T = J T = \alpha_m$ as in Eqs.~(\ref{alpha_m_special_value}), one simply has to replace $N\to M$ in Eq.~(\ref{V_k_res_J}) for~$V_k$, which now becomes   
\be \label{V_k_res_J_M_link}
	V_k = \mu {\mathds 1}  + \nu  \Bigl( {\cal J}_{M-k+1}^{(1,m)} \, {\cal J}_{M-k+2}^{(0,m)} \ldots {\cal J}_{M}^{(0,m)} {\cal J}_{M+1}^{(0,2m)} \ldots {\cal J}_{M+k-1}^{(0,2m)} \, {\cal J}_{M+k}^{(2,2m)} +\text{H.c} \Bigr), \quad 2\leq k \leq K.
\ee
The upper bound on $k$ is given by
\be
	K = \min( M, 2N-M)
\ee 
and it depends on whether the eliminated $M$th link is closer to the left or right boundary of the chain.
Clearly, $V_k$ in Eq.~(\ref{V_k_res_J_M_link}) only changes the internal state of qutrits on the sites~$M-k+1$ and~$M+k$ whereas on the rest of the chain~$V_k$ either produces an extra phase factor or acts trivially.  As a consequence, switching off the $M$th link limits the number of generated entangled qutrit pairs to $K$, which is maximal for~$M=N$. 
Indeed, the entanglement generating protocol also requires some minor modifications. After $k$ periods of the state preparation part of the protocol {\it followed} by $k$ periods of the entanglement generating part of the protocol, the state of the chain is
$
	\left| \psi(k T) \right\rangle = V_1 \ldots V_k \; \otimes_{j=1}^{2N}\,|0\rangle_j,
$
as given by Eq.~(\ref{psi_k_gen}). 
Then, using Eq.~(\ref{V_k_res_J_M_link}) it is straightforward to rewrite $|\psi(kT)\rangle$ as
\be
	 |\psi(kT)\rangle = \bigotimes_{j = 1}^{M-k}\,|0\rangle_j \bigotimes_{j=1}^{k}\,|\Phi (\mu, \nu)\rangle_j \bigotimes_{j = M+k+1}^{2N}\,|0\rangle_j ,
\ee
where $|\Phi (\mu, \nu)\rangle_j$ is given by Eq.~(\ref{Phi_explicit_form}) with the replacement $N \to M$. Finally, taking into account that $k\leq K$, we immediately obtain
\be \label{res_psi_unequal_split}
\begin{aligned}
	& |\psi(K T)\rangle = \bigotimes_{j=1}^{M}\,|\Phi (\mu, \nu)\rangle_j \bigotimes_{j = 2M+1}^{2N}\,|0\rangle_j, && 1\leq M<N, \\
	& |\psi\left( K T\right)\rangle = \bigotimes_{j = 1}^{2(M-N) }\,|0\rangle_j\bigotimes_{j=1}^{2N - M}\,|\Phi (\mu, \nu)\rangle_j , && N < M \leq 2N-1. 
\end{aligned}
\ee
Thus, by switching off the $M$th link one obtains $K$ entangled qutrit pairs, whereas the remaining $2|N-M|$ qutrits are not affected and stay in their initial states. For~$M=N$ one has $K=N$ and Eq.~(\ref{res_psi_unequal_split}) agrees with Eq.~(\ref{final_state}).


\section{Adjoint actions in the Temperley-Lieb algebra} \label{A:TL_adjoints}

In this Appendix we present some useful algebraic relations satisfied by the elements of the Temperley-Lieb algebra $TL_{2M}(\beta)$. From Eqs.~(\ref{TL_rels}) and (\ref{TL_H1_H2}) one obviously has
\be
	\text{ad}_{H_1}H_2 \equiv [H_1,H_2] = \sum_{j=1}^{2M - 2}(-1)^{j} \left[ u_j, u_{j+1} \right]. 
\ee
A little less obvious is the observation that there is a closed form expression for any number of nested commutators:

\be \label{ad_Hm_Hn}
	\text{ad}^k_{H_m}H_n = \bigl[\,\underbrace{H_m, [ H_m, \ldots [H_m}_{k}, H_n]]\bigr] = 
	\begin{cases}
		{\cal F}_{k}^{(m)} ( \beta ), \qquad k = 2 l - 1,\\
		{\cal G}_{k}^{(m)} (\beta), \qquad k = 2 l,
	\end{cases}
\ee
where  $m, n \in \{1,2 \}$, $m \neq n$, and $l$ is a positive integer. Explicitly, the functions ${\cal F}_{k}^{(m)}$ and ${\cal G}_{k}^{(m)}$ read as
\be \label{Fmn_Gmn}
\begin{aligned}
	&{\cal F}_{k}^{(1)}(\beta) = \beta^{k-1} [H_1,H_2] + (2^{k}-2)\beta^{k-2} \, {\cal K}_0, \qquad
	 {\cal F}_{k}^{(2)}(\beta) = - \beta^{k-1} [H_1, H_2] \; + \; (2^{k}-2)\beta^{k-2} {\cal K}_1,\\ 
	\\
	&{\cal G}_{k}^{(1)}(\beta) = - 4\beta^{k-2}\, H_1 +  \beta^{k-2} \,\left( \beta {\cal A} - 2 {\cal R}_0 \right)
	 + (2^k - 2) \beta^{k-2} \, {\cal P}_0 - (2^{k+1} - 8) \beta^{k-3} \, {\cal S}_0, \\ 
	 \\
	&{\cal G}_{k}^{(2)}(\beta) = - 4 \beta^{k-2} \bigl[ H_2 - \frac{1}{2}( u_1 + u_{2M-1}) \bigr] \;+\; \beta^{k-2}\left( \beta {\cal A} - 2 {\cal R}_1 \right)  \;+\; (2^{k}-2)\beta^{k-2} {\cal P}_1
	\;-\; (2^{k+1}-8)\beta^{k-3} {\cal S}_1,
\end{aligned}
\ee
where ${\cal A}$ is given by Eq.~(\ref{calA_operator}) and we introduced the following $k$-independent operators ($n = 0,1$):
\be \label{adHmHn_ops}
\begin{aligned}
	&{\cal K}_{n} =\sum_{j=1}^{M - 2 + n} \left[ u_{ 2j - n } u_{ 2j + 2 - n }, u_{2j + 1 - n} \right] ,  
	&&{\cal R}_{n} = \sum_{j=1}^{M - 2 + n} \left( u_{2j - n} u_{2j+1 - n} u_{2j + 2 - n} + u_{2j + 2 - n} u_{2j+1 - n} u_{2j - n} \right)  , \\
	&{\cal P}_{n} =  \sum_{j=1}^{M - 2 + n} \left\{ u_{ 2j - n } u_{2j + 2 - n}, u_{2j + 1 - n} \right\} ,
	&&{\cal S}_{n} = \sum_{j=1}^{M - 2 + n} u_{2j - n} u_{2j + 2 - n}.
\end{aligned}
\ee
The proof is by induction and is left as an exercise.  One can also check that 
\be \label{ad_Hm_calA}
	\text{ad}^k_{H_m} {\cal A} =\begin{cases}
		 \beta\, {\cal F}_{k}^{(m)}(\beta) \, + \, 2^{k}\beta^{k-1}\,{\cal K}_{m-1},  & k = 2l-1,\\
		 \beta\, {\cal G}_{k}^{(m)}(\beta) \, + \, 2^{k}\beta^{k-2}\,\left( \beta {\cal P}_{m-1} -  2 {\cal S}_{m-1} \right), 	& k=2l.
		 \end{cases}
\ee

Eqs. (\ref{ad_Hm_Hn})-(\ref{ad_Hm_calA}) allow us to calculate various adjoint actions in a closed form. Let us consider
\be \label{Ad_e_Hm_Hn}
	\tilde H_n(s) \equiv e^{ s H_m} H_n\, e^{ -s H_m } = e^{s \, \text{ad}_{H_m}} H_n = \sum_{k=0}^{+\infty} \frac{s^k}{k!}\text{ad}^k_{H_m} H_n = H_n + \sum_{l=1}^{+\infty} \frac{s^{2l-1}}{(2l-1)!}{\cal F}_{2l-1}^{(m)}(\beta) + \sum_{l=1}^{+\infty} \frac{s^{2l}}{(2l)!}{\cal G}_{2l}^{(m)}(\beta).
\ee
Then, using Eq. (\ref{Fmn_Gmn}), we immediately obtain
\be \label{tildeH_2}
\begin{aligned}
	\tilde H_2(s) &= e^{ s H_1} H_2\, e^{ -s H_1 } = H_2 -\frac{8}{\beta^2}\sinh^2 \frac{s \beta}{2}\, H_1 + \frac{1}{\beta}\sinh s \beta \, \left[ H_1, H_2 \right] + \frac{4}{\beta^2} \sinh s \beta \, \sinh^2 \frac{s \beta}{2} \, {\cal K}_0\\
	&+ \frac{2}{\beta^2} \sinh^2 \frac{s\beta}{2} \left( \beta {\cal A} - 2 {\cal R}_0 \right) 
	+\frac{4}{\beta^2}\cosh s \beta \, \sinh^2\frac{s \beta}{2} \,{\cal P}_0 - \frac{16}{\beta^3}\sinh^4 \frac{s\beta}{2}\, {\cal S}_0.
\end{aligned}
\ee
Similarly, for~$\tilde H_1(s)$ one has the following expression:
\be \label{tildeH_1}
\begin{aligned}
	\tilde H_1(s) &= e^{ s H_2} H_1\, e^{ -s H_2 } = H_1 -\frac{8}{\beta^2}\sinh^2 \frac{s \beta}{2}\, \bigl[ H_2 - \frac{1}{2}( u_1 + u_{2M-1}) \bigr] - \frac{1}{\beta}\sinh s \beta \, \left[ H_1, H_2 \right] \\
	&+ \frac{4}{\beta^2} \sinh s \beta \, \sinh^2 \frac{s \beta}{2} \, {\cal K}_1 + \frac{2}{\beta^2} \sinh^2 \frac{s\beta}{2} \Bigl( \beta {\cal A} - 2 {\cal R}_1 \Bigr)
	+\frac{4}{\beta^2}\cosh s \beta \, \sinh^2\frac{s \beta}{2} \,{\cal P}_1 - \frac{16}{\beta^3}\sinh^4 \frac{s\beta}{2} \, {\cal S}_1.
\end{aligned}
\ee
In the same way one can calculate
\be \label{Ad_e_Hm_ad_Hm_Hn}
	e^{s H_m} [H_m, H_n] e^{- s H_m} = \sum_{k=0}^{+\infty} \frac{s^k}{k!} \text{ad}_{H_m}^{k+1} H_n = [H_m, H_n] + \sum_{l=1}^{+\infty} \frac{s^{2l-1}}{(2l-1)!}{\cal G}_{2l}^{(m)}(\beta) + \sum_{l=1}^{+\infty} \frac{s^{2l}}{(2l)!}{\cal F}_{2l+1}^{(m)}(\beta), 
\ee
which yields
\be \label{Ad_e_Hm_ad_Hm_Hn_1_2}
\begin{aligned}
	e^{s H_1} &  [H_1, H_2] e^{- s H_1}  = \cosh s \beta \, [H_1, H_2]  - \frac{4}{\beta} \sinh s \beta H_1  + \frac{4}{\beta} \sinh\frac{3 s \beta}{2} \sinh \frac{s \beta}{2} {\cal K}_0 \\ 
	&+ \frac{1}{\beta} \sinh s \beta \left( \beta {\cal A} - 2 {\cal R}_0 \right)  + \frac{4}{\beta} \cosh\frac{3 s \beta}{2} \sinh \frac{s \beta}{2} {\cal P}_0  - \frac{16}{\beta^2} \sinh s \beta \sinh^2\frac{s \beta}{2} {\cal S}_0,\\
	\\
	e^{s H_2} &  [H_2, H_1] e^{- s H_2}  =\cosh s \beta \, [H_2,H_1]  - \frac{4}{\beta} \sinh s \beta \bigl[ H_2 - \frac{1}{2}(u_1 + u_{2M-1}) \bigr] + \frac{4}{\beta} \sinh\frac{3 s \beta}{2} \sinh \frac{s \beta}{2} {\cal K}_1 \\ 
	&+ \frac{1}{\beta} \sinh s \beta \left( \beta {\cal A} - 2 {\cal R}_1 \right)  + \frac{4}{\beta} \cosh\frac{3 s \beta}{2} \sinh \frac{s \beta}{2} {\cal P}_1  - \frac{16}{\beta^2} \sinh s \beta \sinh^2 \frac{s \beta}{2} {\cal S}_1.
\end{aligned}
\ee
Proceeding as in Eq. (\ref{Ad_e_Hm_Hn}), it is also easy to show that
\be \label{Ad_e_Hm_calA}
\begin{aligned}
	&e^{s H_{1}} {\cal A} e^{- s H_{1}} = {\cal A} + \beta \left( \tilde H_2(s) - H_2 \right) + \frac{1}{\beta}\sinh 2s\beta \,{\cal K}_{0} + \frac{2}{\beta^2}\sinh^2s\beta\, \left(\beta {\cal P}_{0} - 2 {\cal S}_{0} \right), \\
	&e^{s H_{2}} {\cal A} e^{- s H_{2}} = {\cal A} + \beta \left( \tilde H_1(s) - H_1 \right) + \frac{1}{\beta}\sinh 2s\beta \,{\cal K}_{1} + \frac{2}{\beta^2}\sinh^2s\beta\, \left(\beta {\cal P}_{1} - 2 {\cal S}_{1} \right). \\
\end{aligned}
\ee

It is now straightforward to check whether the ansatz~(\ref{Q1_ansatz}) for the first conserved charge satisfies the integrability condition~(\ref{U_F_integrability_condition}). Using Eqs.~(\ref{tildeH_2}), (\ref{tildeH_1}), (\ref{Ad_e_Hm_ad_Hm_Hn_1_2}), and~(\ref{Ad_e_Hm_calA}), one finds that~$[U_F, Q_1] = 0$ only if~$T_1 = T_2$ and the coefficients~$a_k$ are given by Eq.~(\ref{Q1_ansatz_coeffs}).

\end{widetext}


\begin{thebibliography}{99}

\bibitem{Lukin2017}
H. Bernien, S. Schwartz, A. Keesling, H. Levine, A. Omran, H. Pichler, S. Choi, A.S. Zibrov, M. Endres, M. Greiner, V. Vuleti{\'c}, and M.D. Lukin,
Probing many-body dynamics on a 51-atom quantum simulator,
{\href{http://dx.doi.org/10.1038/nature24622}{Nature (London) {\bf 551}, 579 (2017)}}.

\bibitem{Monroe2017} 
J. Zhang, G. Pagano, P. W. Hess, A. Kyprianidis, P. Becker, H. Kaplan, A.V. Gorshkov, Z.-X. Gong, and C. Monroe,
Observation of a many-body dynamical phase transition with a 53-qubit quantum simulator,
{\href{http://dx.doi.org/10.1038/nature24654}{Nature (London) {\bf 551}, 601 (2017)}}.

\bibitem{Martinis2018}
C. Neill, P. Roushan, K. Kechedzhi, S. Boixo, S.V. Isakov, V. Smelyanskiy, R. Barends, B. Burkett, Y. Chen, Z. Chen, 
B. Chiaro, A. Dunsworth, A. Fowler, B. Foxen, R. Graff, E. Jeffrey, J. Kelly, E. Lucero, A. Megrant, J. Mutus, M. Neeley, 
C. Quintana, D. Sank, A. Vainsencher, J. Wenner, T.C. White, H. Neven, and J.M. Martinis,
A blueprint for demonstrating quantum supremacy with superconducting qubits,
{\href{http://dx.doi.org/10.1126/science.aao4309}{Science {\bf 360}, 195 (2018)}}.

\bibitem{Blatt2018}
N. Friis, O. Marty, C. Maier, C. Hempel, M. Holz{\"a}pfel, P. Jurcevic, M.B. Plenio, M. Huber, C. Roos, R. Blatt, and B. Lanyon,
Observation of entangled states of a fully controlled 20-qubit system,
{\href{http://dx.doi.org/10.1103/PhysRevX.8.021012}{Phys. Rev. X {\bf 8}, 021012 (2018)}}.

\bibitem{Trotzky2012}
S. Trotzky, Y.-A. Chen, A. Flesch, I.P. McCulloch, I. Schollw\"{o}ck, J. Eisert, and I. Bloch,
Probing the relaxation towards equilibrium in an isolated strongly correlated one-dimensional Bose gas,
{\href{http://dx.doi.org/10.1038/nphys2232}{Nat. Phys. {\bf 8}, 325 (2012)}}.

\bibitem{Mazurenko2017}
A. Mazurenko, C.S. Chiu, G. Ji, M.F. Parsons, M. Kan{\'a}sz-Nagy, R. Schmidt, F. Grusdt, E. Demler, D. Greif, and M. Greiner,
A cold-atom Fermi-Hubbard antiferromagnet,
{\href{http://dx.doi.org/10.1038/nature22362}{Nature (London) {\bf 545}, 462 (2017)}}.

\bibitem{Lukin2019}
A. Keesling, A. Omran, H. Levine, H. Bernien, H. Pichler, S. Choi, R. Samajdar, S. Schwartz, P. Silvi, S. Sachdev, P. Zoller, M. Endres, M. Greiner, V. Vuleti{\'c}, and M.D. Lukin,
Quantum Kibble--Zurek mechanism and critical dynamics on a programmable Rydberg simulator,
\href{https://doi.org/10.1038/s41586-019-1070-1}{Nature (London) {\bf 568}, 207 (2019)}.

\bibitem{Sadchev2018}
R. Samajdar, S. Choi, H. Pichler, M. D. Lukin, and S Sachdev,
Numerical study of the chiral $\mathbb{Z}_3$ quantum phase transition in one spatial dimension,
{\href{http://dx.doi.org/10.1103/PhysRevA.98.023614}{Phys. Rev. A {\bf 98}, 023614 (2018)}}.

\bibitem{Gorshkov2020}
F. Liu, S. Whitsitt, P. Bienias, R. Lundgren, and A. V. Gorshkov,
Realizing and probing baryonic excitations in Rydberg atom arrays,
\href{https://arxiv.org/abs/2007.07258}{arXiv:2007.07258}.

\bibitem{integrability}
P. P. Martin, {\it Potts Models and Related Problems in Statistical Mechanics} (World
Scientific, 1991)

\bibitem{topological}
A. Hutter, J. R. Wootton, and D. Loss, Parafermions in a Kagome lattice of qubits for topological quantum computation, \href{https://doi.org/10.1103/PhysRevX.5.041040}{Phys. Rev. X {\bf 5}, 041040 (2015)}.

\bibitem{Fendley2016}
J. Alicea and P. Fendley,
Topological phases with parafermions: theory and blueprints,
\href{https://doi.org/10.1146/annurev-conmatphys-031115-011336}{Annu. Rev. Condens. Matter Phys. {\bf 7}, 119 (2016)}.

\bibitem{HutterLoss2016}
A. Hutter and D. Loss,
Quantum computing with parafermions,
\href{https://doi.org/10.1103/PhysRevB.93.125105}{Phys. Rev. B {\bf 93}, 125105 (2016)}.

\bibitem{White2009}
B.P. Lanyon, M. Barbieri, M.P. Almeida, T. Jennewein, T.C. Ralph, K.J. Resch, G.J. Pryde, J.L. O'Brien, A. Gilchrist, and A.G. White,
Simplifying quantum logic using higher-dimensional Hilbert spaces,
{\href{http://dx.doi.org/10.1038/nphys1150}{Nat. Phys. {\bf 5}, 134 (2009)}}.

\bibitem{Wallraff2012}
A. Fedorov, L. Steffen, M. Baur, M.P. da Silva, and A. Wallraff,
Implementation of a Toffoli gate with superconducting circuits,
{\href{http://dx.doi.org/10.1038/nature10713}{Nature (London) {\bf 481}, 170 (2012)}}.

\bibitem{Kiktenko2020}
E.O. Kiktenko, A.S. Nikolaeva, Peng Xu, G.V. Shlyapnikov, and A.K. Fedorov, 
Scalable quantum computing with qudits on a graph, 
\href{https://doi.org/10.1103/PhysRevA.101.022304}{Phys. Rev. A {\bf 101}, 022304 (2020)}. 

\bibitem{Gokhale2019}
P. Gokhale, J. M. Baker, C. Duckering, N. C. Brown, K. R. Brown, and F. T. Chong, 
P. Gokhale, J. M. Baker, C. Duckering, N. C. Brown, K. R. Brown, and F. T. Chong, Asymptotic improvements to quantum circuits via qutrits, in {\it Proceedings of the 46th International Symposium on Computer Architecture} (Association for Computing Machinery, New York, 2019).

\bibitem{Anderson1987}
P. W. Anderson, 
The resonating valence bond state in La$_2$CuO$_4$ and superconductivity,
\href{httpsL//dx.doi.org/10.1126/science.235.4793.1196}{Science {\bf 235}, 1196 (1987)}.


\bibitem{Gritsev2008}
P. Barmettler, A.\,M. Rey, E. Demler, M.\,D. Lukin, I. Bloch, and V. Gritsev, 
Quantum many-body dynamics of coupled double-well superlattices, 
\href{https://doi.org/10.1103/PhysRevA.78.012330}{Phys. Rev. A {\bf 78}, 012330 (2008)}. 

 
 \bibitem{Gritsev2017}
 V. Gritsev and A. Polkovnikov, Integrable Floquet dynamics, \href{https://doi.org/10.21468/SciPostPhys.2.3.021}{SciPost Phys. {\bf 2}, 021 (2017)}.
 
\bibitem{Lorenzo2017}
S. Lorenzo, J. Marino, F. Plastina, G. M. Palma, and T. J. G. Apollaro, Quantum Critical Scaling under Periodic Driving \href{https://doi.org/10.1038/s41598-017-06025-1}{Sci. Rep. {\bf 7}, 5672 (2017)}.
 
 \bibitem{Bertini2019a}
B. Bertini, P. Kos, and T. Prosen, Exact correlation functions for dual-unitary lattice models in 1 + 1 Dimensions, \href{https://doi.org/10.1103/PhysRevLett.123.210601}{Phys. Rev. Lett. {\bf 123}, 210601 (2019)}.

\bibitem{Bertini2019b}
B. Bertini, P. Kos, and T. Prosen, Entanglement spreading in a minimal model of maximal many-body quantum chaos, \href{https://doi.org/10.1103/PhysRevX.9.021033}{Phys. Rev. X {\bf 9}, 021033 (2019)}.

\bibitem{Bertini2020a}
B. Bertini, P. Kos, and T. Prosen, Operator entanglement in local quantum circuits I: Chaotic dual-unitary circuits, \href{https://doi.org/10.21468/SciPostPhys.8.4.067}{SciPost Phys. {\bf 8}, 067 (2020)}.

\bibitem{Bertini2020b}
B. Bertini, P. Kos, and T. Prosen, Operator entanglement in local quantum circuits II: solitons in chains of qubits, \href{https://doi.org/10.21468/SciPostPhys.8.4.068}{SciPost Phys. {\bf 8}, 068 (2020)}.

\bibitem{Fan2020}
R. Fan, Y. Gu , A. Vishwanath, and X. Wen, Emergent spatial structure and sntanglement localization in Floquet conformal field theory, \href{https://doi.org/10.1103/PhysRevX.10.031036}{Phys. Rev X {\bf 10}, 031036 (2020)}.

\bibitem{Klobas2020}
 K. Klobas, B. Bertini, and L. Piroli, Exact thermalization dynamics in the ``Rule 54'' Quantum Cellular Automaton, \href{https://doi.org/10.1103/PhysRevLett.126.160602}{Phys. Rev. Lett. {\bf 126}, 160602 (2021)}.
 
 \bibitem{Maskara2021}
 N. Maskara, A. A. Michailidis, W. W. Ho, D. Bluvstein, S. Choi, M. D. Lukin, and M. Serbyn, Discrete time-crystalline order enabled by quantum many-body scars: entanglement steering via periodic driving, \href{https://doi.org/10.1103/PhysRevLett.127.090602}{Phys. Rev. Lett. {\bf 127}, 090602 (2021)}.


\bibitem{Bluvstein2021}
D. Bluvstein, A. Omran, H. Levine, A. Keesling, G. Semeghini, S. Ebadi, T. T. Wang, A. A. Michailidis, N. Maskara, W. W. Ho, S. Choi, M. Serbyn, M. Greiner, V. Vuletic, and M. D. Lukin,
Controlling many-body dynamics with driven quantum scars in Rydberg atom arrays,
{\href{http://dx.doi.org/10.1126/science.abg2530}{Science {\bf 371}, 1355 (2021)}}.

\bibitem{Lakshminarayan2005}
A. Lakshminarayan and V. Subrahmanyam, 
Multipartite entanglement in a one-dimensional time-dependent Ising model, 
\href{https://doi.org/10.1103/PhysRevA.71.062334}{Phys. Rev. A {\bf 71}, 062334 (2005)}.

\bibitem{Mishra2015}
S.\,K. Mishra, A. Lakshminarayan, and V. Subrahmanyam, 
Protocol using kicked Ising dynamics for generating states with maximal
multipartite entanglement, \href{https://doi.org/10.1103/PhysRevA.91.022318}{Phys. Rev. A {\bf 91}, 022318 (2015)}.

\bibitem{Pal2018}
R. Pal and A. Lakshminarayan, 
Entangling power of time-evolution operators in integrable and nonintegrable many-body systems, 
\href{https://doi.org/10.1103/PhysRevB.98.174304}{Phys. Rev. B {\bf 98}, 174304 (2018)}.
 
\bibitem{Naik2019}
G.\,K. Naik, R. Singh, and S.\,K. Mishra, 
Controlled generation of genuine multipartite entanglement in Floquet Ising spin models, 
\href{https://doi.org/10.1103/PhysRevA.99.032321}{Phys. Rev. A {\bf 99}, 032321 (2019)}.

\bibitem{Lazarides2014}
A. Lazarides, A. Das, and R. Moessner, Equilibrium states of generic quantum systems subject to periodic driving, \href{https://doi.org/10.1103/PhysRevE.90.012110}{Phys. Rev. E {\bf 90}, 012110 (2014)}.

\bibitem{DAlessio2014}
L. D'Alessio and M. Rigol, Long-time behavior of isolated periodically driven interacting lattice systems, \href{https://doi.org/10.1103/PhysRevX.4.041048}{Phys. Rev. X {\bf 4}, 041048 (2014)}.

\bibitem{Ponte2015}
P. Ponte, A. Chandran, Z. Papic, and D. A. Abanin, Periodically driven ergodic and many-body localized quantum systems, \href{https://doi.org/10.1016/j.aop.2014.11.008}{Ann. Phys. {\bf 353}, 196 (2015)}

\bibitem{Srednicki1994}
M. Srednicki, Chaos and quantum thermalization, \href{https: //link.aps.org/doi/10.1103/PhysRevE.50.888}{Phys. Rev. {\bf E} 50, 888 (1994)}.

\bibitem{Rigol2016}
M. Rigol, V. Dunjko, and M. Olshanii, Thermalization and its mechanism for generic isolated quantum systems, \href{http://dx.doi.org/10.1038/nature06838}{Nature {\bf 452}, 854 (2016)}.

\bibitem{boundaries_note}
	The choice of boundary conditions mostly affects the results of Sec.~\ref{S:integrability} because of the connection with the Temperley-Lieb algebra~$TL_{L}(\beta)$, which requires open boundary conditions. However, it is quite simple to extend the results of Sec.~\ref{S:integrability} to the case of periodic boundary conditions by using the periodic version of the Temperley-Lieb algebra~$pTL_{L}(\beta)$. 

\bibitem{gen_Pauli_Y_op}
Note that one can define an operator~$Y_j = X_j Z_j$, which then satisfies the relations~$X_j Y_j = \omega Y_j X_j$ and $Y_j Z_j = \omega Z_j Y_j$, similar to those for~$X_j$ and~$Z_j$ in Eq.~(\ref{Z_X_props}). However, it is easy to see that neither $\{X_j, Y_j, Z_j \}$ nor $\{ X_j^m, Y_j^p, Z_j^q  \}$ with $m,p,q \in \{ 0, 1,2 \}$ form a group or a Lie algebra. For this reason it is more convenietn to work with the basis of~${}\mathfrak sl(3,{\mathbb C})$ in Eq.~(\ref{Z_X_to_J}).

\bibitem{Patera87} J. Patera and H. Zassenhaus, The Pauli matrices in $n$ dimensions and finest gradings of simple Lie algebras of type $A_{n-1}$, \href{http://dx.doi.org/10.1063/1.528006}{J. Math. Phys. {\bf 29}, 665 (1988)}.

\bibitem{Fairliea89} D.\,B. Fairliea, C.\,K. Zachos, Infinite-dimensional algebras, sine brackets, and SU($\infty$), \href{https://doi.org/10.1016/0370-2693(89)91057-5}{Phys. Lett. A {\bf 224}, 101 (1989)}. 

\bibitem{J_ops_definition_note} Note that the definition of~${\cal J}_j^{\boldsymbol{m}}$ in Eq.~(\ref{Z_X_to_J}) is slightly different from the conventional one. The latter reads~$\tilde {\cal J}_j^{\boldsymbol{m}} = \omega^{m_1 m_2 /2} Z_j^{m_1} X^{m_2}_j$, see~Ref.~\cite{Fairliea89}, which also modifies Eqs.~(\ref{J_J_product}) and (\ref{J_comm_rel}). We stress that our definition is consistent, i.e. starting from Eq.~(\ref{Z_X_to_J}) one can obtain~Eqs.~(\ref{J_J_product}) and (\ref{J_comm_rel}), at least for $n=3$. In addition, the advantage of~Eq.~(\ref{Z_X_to_J}) is that the indices~$m_1$ and~$m_2$ in~${\cal J}_j^{(m_1, m_2)}$ are automatically evaluated $\text{mod}\;3$.

\bibitem{superintegrability} 
G. von Gehlen and V. Rittenberg, $Z_n$-symmetric quantum chains with an infinite set of conserved charges and $Z_n$ zero modes, \href{https://doi.org/10.1016/0550-3213(85)90350-5}{Nucl. Phys. B {\bf 257}, 351 (1985)}.

\bibitem{Bukov2015}
M. Bukov, L. D'Alessio, and A. Polkovnikov, Universal high-frequency behavior of periodically driven systems: from dynamical stabilization to Floquet engineering, \href{http://dx.doi.org/10.1080/00018732.2015.1055918}{Advances in Physics, {\bf 64}, 139 (2015)}.

\bibitem{Arze2020}
S. E. T. Arze, P. W. Claeys, I. P. Castillo, and J.-S. Caux, Out-of-equilibrium phase transitions induced by Floquet resonances in a periodically quench-driven XY spin chain, \href{https://doi.org/10.21468/SciPostPhysCore.3.1.001}{SciPost Phys. Core {\bf 3}, 001 (2020)}.

\bibitem{note_unequal_split}
The case of switching off non-central links is discussed in Appendix~\ref{A:Floquet_protocol_unequal_split}, where we present the resulting entangled states.

\bibitem{note_initial_state}
	The entanglement generating protocol also works for any initial state that is an eigenstate of the operator~$\prod_{j=1}^{2N} Z_j$, with~$Z_j$ given by Eq.~(\ref{Z_X_matrix_rep}). In other words $\ket{\psi(0)}$ must be a product state although it can be more complicated than the polarized state~(\ref{polarized_init_state}).

\bibitem{Sych2009}
D. Sych and G. Leuchs, A complete basis of generalized Bell states, \href{https://doi.org/10.1088/1367-2630/11/1/013006}{New J. Phys. {\bf 11}, 013006 (2009)}.

\bibitem{Ridout2014}
D. Ridout and Y. Saint-Aubin, Standard modules, induction and the structure of the Temperley-Lieb algebra, \href{https://doi.org/10.4310/ATMP.2014.v18.n5.a1}{Adv. Theor. Math. Phys. {\bf 18}, 957 (2014)}.

\bibitem{TL_Groebner_basis}
J.-Y. Lee, D.-I. Lee, and S. Kim, Gr\"obner-Shirshov bases for Temperley-Lieb algebras of complex reflection groups, \href{https://doi.org/10.3390/sym10100438}{Symmetry {\bf 10}, 438 (2018)}.

\bibitem{Nichols2006}
A. Nichols, The Temperley-Lieb algebra and its generalizations in the Potts and XXZ models, \href{https://doi.org/10.1088/1742-5468/2006/01/P01003}{J. Stat. Mech. {\bf 2006}, P01003 (2006)}.

\bibitem{Q1_note}
Note that the term in Eq.~(\ref{TL_H_1_H_2_comm}) has the form of the next correction (in terms of the high-frequency expansion) to the Floquet Hamiltonian. Note also that the alternating factor $(-1)^{j}$ was missing in the corresponding expression in Ref.~\cite{Gritsev2017}.

\bibitem{Grabowski1996}
M.\,P. Grabowski and P. Mathieu, The structure of conserved charges in open spin chains, 
\href{https://doi.org/10.1088/0305-4470/29/23/024}{J.Phys. A {\bf 29}, 7635 (1996)}.
 
 \bibitem{Ungar1982} A. Ungar, Generalized hyperbolic functions. \href{https://doi.org/10.2307/2975654}{Amer. Math. Monthly {\bf 89}, 688 (1982)}. 
 
 \bibitem{Muldoon1996} M.E. Muldoon and A. Ungar, Beyond sin and cos, \href{https://doi.org/10.1080/0025570X.1996.11996374}{Math. Mag. {\bf 69}, 3 (1996)}.
 
 \bibitem{Muldoon2005} M.E. Muldoon, Generalized hyperbolic functions, circulant matrices and functional equations, \href{https://doi.org/10.1016/j.laa.2005.04.011}{Linear Algebra Appl. {\bf 406}, 272 (2005)}.
 



 
 

\end{thebibliography}
\end{document}